\documentclass[a4paper,fleqn]{cas-dc}

\usepackage[authoryear]{natbib}
\usepackage{xcolor}
\usepackage[utf8]{inputenc}
\usepackage{multirow}
\usepackage{longtable}
\usepackage{booktabs}
\usepackage{tabularx}
\usepackage{graphicx}
\usepackage{amssymb,amsmath}
\usepackage{hyperref}
\usepackage{pdflscape}
\usepackage{mathrsfs}
\usepackage{array}
\usepackage{pifont}
\usepackage{rotating}
\usepackage{ulem}
\usepackage{aas_macros}

\begin{document}

\let\oldAA\AA
\renewcommand{\AA}{\text{\normalfont\oldAA}}

\shorttitle{Supermassive Black Holes Mass}

\shortauthors{Lu et al.}

\title [mode = title]{287,872 Supermassive Black Holes Masses: Deep Learning Approaching Reverberation Mapping Accuracy}

\author[1,2]{Yuhao Lu\footnote{ORCID: 0009-0002-7655-6737}}

\affiliation[1]{organization={School of Computer Science, University of South China},
            addressline={Hengyang}, 
            postcode={421001}, 
            country={China}}
            
\affiliation[2]{organization={ICRANet-AI},
            addressline={Brickell Avenue 701}, 
            city={Miami},
            postcode={FL 33131}, 
            country={USA}}

\author[3]{HengJian SiTu\footnote{ORCID: 0009-0005-4130-6107}}

\affiliation[3]{organization={School of Mathematics and Physics, University of South China},
            addressline={Hengyang}, 
            postcode={421001}, 
            country={China}}

\author[3]{Jie Li\footnote{ORCID: 0009-0005-2280-1719}}

\author[3,2]{Yixuan Li\footnote{ORCID: 0009-0002-0541-7182}}

\author[1,2,4]{Yang Liu\footnote{ORCID: 0009-0002-9188-5029}}

\affiliation[4]{organization={Department of Physics E. Pancini, University Federico II},
            addressline={Naples}, 
            postcode={80126}, 
            country={Italy}}

\author[1,3,5]{Wenbin Lin\footnote{ORCID: 0000-0002-4282-066X}}

\affiliation[5]{organization={ICRANet},
            addressline={Piazza della Repubblica 10}, 
            city={Pescara},
            postcode={65122}, 
            country={Italy}}
\ead{lwb@usc.edu.cn}

\author[6,2,5,7]{Yu Wang\footnote{ORCID: 0000-0001-7959-3387}\cormark[1]}

\ead{yu.wang@icranet.org}

\affiliation[6]{organization={ICRA and Dipartimento di Fisica, Sapienza Universit\`a di Roma},
            addressline={P.le Aldo Moro 5}, 
            city={Rome},
            postcode={00185}, 
            country={Italy}}

\affiliation[7]{organization={INAF -- Osservatorio Astronomico d'Abruzzo},
            addressline={Via M. Maggini snc}, 
            city={Teramo},
            postcode={I-64100}, 
            country={Italy}}

\cortext[1]{Corresponding author}

\begin{abstract}
We present a population-scale catalogue of 287,872 supermassive black hole masses with high accuracy. Using a deep encoder-decoder network trained on optical spectra with reverberation-mapping (RM) based labels of 849 quasars and applied to all SDSS quasars up to $z=4$, our method achieves a root-mean-square error of $0.058$\,dex, a relative uncertainty of $\approx 14\%$, and coefficient of determination $R^{2}\approx0.91$ with respect to RM-based masses, far surpassing traditional single-line virial estimators. Notably, the high accuracy is maintained for both low ($<10^{7.5}\,M_\odot$) and high ($>10^{9}\,M_\odot$) mass quasars, where empirical relations are unreliable.
\end{abstract}


\maketitle

\section{Introduction}
\label{sec:intro}

Supermassive black holes (SMBHs) with masses spanning from roughly $10^5\,M_\odot$ to $10^{10}\,M_\odot$ are commonly observed at the centers of most massive galaxies \citep{1995ARA&A..33..581K,2005SSRv..116..523F,2013ARA&A..51..511K}. Recent breakthroughs, particularly the Event Horizon Telescope's imaging of the SMBH at the core of the elliptical galaxy M\,87, have provided unprecedented direct observational evidence \citep{2019ApJ...875L...1E}. It is now firmly established that SMBH masses are strongly correlated with the characteristics of their host galaxies, including bulge mass, stellar velocity dispersion, surface brightness, and luminosity \citep{2000ApJ...539L...9F,2001ApJ...547..140M,2004ApJ...604L..89H,2016ApJ...818...47S}. These correlations appear to persist across both local and high-redshift galaxies \citep{2013ApJ...764..151G,2013ApJ...767...13S,2019PASJ...71..111I}, suggesting a fundamental co-evolutionary link despite the vast difference in physical scales between SMBHs and their hosts \citep{2008ApJS..175..356H,2010ApJ...711..284S,2019PASJ...71..111I}.

Nevertheless, significant challenges remain, particularly in understanding how such massive black holes could have formed within the universe's first billion years \citep{2015Natur.518..512W,2020ARA&A..58...27I}. Current models suggest that SMBHs grow predominantly through gas accretion and galaxy mergers, releasing substantial energy that profoundly affects host galaxy evolution \citep{2012NewAR..56...93A,2007ApJ...665.1038C,2007MNRAS.380..877S}.

Observationally, SMBHs manifest as active galactic nuclei (AGNs) or quasars, whose extreme luminosities offer key insights into accretion processes and black hole growth \citep{1982MNRAS.200..115S}. However, directly measuring SMBH masses remains challenging. Spectroscopic techniques, though widely used, are labor-intensive and have yielded only about one million estimates over the past two decades \citep{2011ApJS..194...45S,2013ApJ...764...45K}. The advent of large-scale surveys such as the Vera C. Rubin Observatory's Legacy Survey of Space and Time (LSST), expected to detect nearly $10^8$ quasars \citep{2019ApJ...873..111I}, will require far more efficient and scalable mass estimation methods. Alternative approaches based on variability measurements in optical and X-ray bands show promise \citep{2006Natur.444..730M,2021Sci...373..789B}, but are complicated by nonlinear physical dependencies and the massive data volumes involved. Meanwhile, direct dynamical measurements of SMBH masses remain limited to only a small sample of nearby galaxies \citep{2011ApJ...727...20K,2013ARA&A..51..511K,2013ApJ...764..184M,2016MNRAS.460.3119S,2019MNRAS.485.1278S}.

While classical machine learning techniques, such as symbolic regression, random forests, and photometric regressions, have contributed to early progress in black hole mass estimation by extending traditional scaling relations \citep{2023arXiv231019406J, 2022RAA....22h5014H}, their performance remains fundamentally constrained by shallow architectures and limited feature representations. In contrast, deep learning approaches have shown greater promise in capturing the non-linear dependencies inherent in high-dimensional astrophysical data. For instance, variability-based models such as AGNet achieve a scatter of approximately 0.37 dex with respect to virial mass estimates derived from single-epoch spectra \citep{2023MNRAS.518.4921L}. On the other hand, models trained on X-ray reverberation-mapped AGNs, where more physically grounded mass measurements are available, have demonstrated predictive accuracies within 2-5\% of the reference values \citep{2022MNRAS.513..648C}. Although these performance metrics are not directly comparable due to differing calibration standards, they collectively highlight the advantages of deep learning over traditional approaches.

To further advance black hole mass estimation, we propose a deep learning-based framework, QuasarSpecNet, which integrates a convolutional neural network (CNN) and a transposed convolutional network (DeconvNet). The model utilizes an encoding-decoding architecture to process spectral data, and performs mass prediction through representations learned in a latent space. Through joint training, the model not only enables efficient reconstruction of the input spectra but also allows accurate prediction of black hole masses based on latent features. The architecture is designed to leverage the hierarchical feature extraction capabilities of deep convolutional layers, while incorporating skip connections to retain fine-grained spectral details during decoding. This design improves predictive precision and more effectively preserves physical information. Additionally, it introduces a novel approach for estimating black hole masses and demonstrates superior performance in processing and analyzing spectroscopic data. The method achieves a root-mean-square error of 0.05 dex in $\log(M_\mathrm{BH})$.

A central aspect of this work is that the network is trained on SDSS-RM spectra with reverberation-mapped black hole masses as labels, which provide a relatively small (849 quasars) but physically accurate sample. After training, the model is applied to an extended SDSS sample of 287,872 quasars spanning $0 \lesssim z \lesssim 4$, producing mass estimates across $\log_{10}(M_{\rm BH}/M_{\odot}) \approx 6$--$10$. The combination of RM-based training and large-scale application is the main advance of this study. It provides a uniform mass catalog large enough to trace black hole demographics over cosmic time, while combining the accuracy of RM calibration with the statistical power of survey-scale quasar populations.

This article is structured as follows: Section \ref{sec:data} describes the dataset and preprocessing steps; Section \ref{sec:model} details the proposed methodology; Section \ref{sec:evaluation} presents model evaluation metrics; Section \ref{sec:287,872 Black Hole Mass} applies the trained model to 287,872 SDSS quasars and presents the resulting mass catalog; Section \ref{sec:discussion} discusses the findings; finally, Section 7 provides conclusions and acknowledgments.

\section{Dataset}
\label{sec:data}

\subsection{Extraction of Observational Parameters}

For the SDSS-RM sample, each quasar was observed repeatedly over multiple epochs, producing time-series spectra that capture the variability of both continuum and broad emission lines. Reverberation mapping relies on measuring the delayed response of lines such as H$\alpha$, H$\beta$, Mg \textsc{ii}, and C\textsc{iv} to continuum changes.  Each emission line is recorded across many epochs, enabling variability studies and lag determinations. As a result, the dataset contains both single-epoch observational parameters and multi-epoch spectral sequences.

Observational metadata are extracted from the headers of the raw spectral files. Key parameters including the spectroscopic plate number (PLATE), observation date (MJD), fiber identifier (FIBERID), mean observation epoch (Mean\_MJD), redshift (ZFINAL), and observation epoch index (EPOCH) are extracted using regular expressions. These parameters were organized into structured records to facilitate sample selection, redshift dependent analyses, and cross-validation of spectral line measurements. 
Table~\ref{tab:dataset_fields} summarizes these observational parameters. 

The SDSS-RM sample includes 849 unique sources with 75{,}915 spectra, spanning redshifts from $z = 0.12$ to $z = 4.33$ (mean $z = 1.64$, median $z = 1.61$). Only three sources (270 spectra) lie at $z \ge 4$, indicating that the sample is dominated by low- to intermediate-redshift AGNs.

\subsection{Data Preprocessing and Standardization}
\label{subsec:tables}

To provide consistent and physically meaningful input spectra for the deep learning analysis, we preprocess quasar spectra obtained from the SDSS-RM project \citep{2011ApJS..194...45S,2020ApJS..250....8L,shen2024sloan}. This preprocessing procedure involves extraction of key observational parameters, resampling of spectra to a uniform wavelength grid, exclusion of invalid or corrupted spectra, spectral normalization, and the compilation of a standardized dataset suitable for both training and evaluation of our model. Each preprocessing step is described in detail below.

\subsection{Wavelength Resampling}
The original spectral flux data points are sampled unevenly and lie on a non-uniform wavelength grid. To standardize the model input, the wavelengths were first transformed into logarithmic space and then interpolated onto a uniform grid of length 4618 (denoted as \texttt{Wave1}):
\begin{equation}
                   \lambda_i^\ast = \lambda_{\min} + i \cdot \Delta\lambda,
\end{equation}
\begin{equation}
                   \quad \Delta\lambda = \frac{\lambda_{\max} - \lambda_{\min}}{4617},
\end{equation}
where $i$ ranges from 0 to 4617, and the wavelength range corresponds to 360--1032.5\,nm (vacuum wavelength). Cubic spline interpolation was employed as the primary method, with linear extrapolation used at the boundaries. The resulting flux values were stored as \texttt{Flux1}, with units preserved as $10^{-17}\,\mathrm{erg\,s^{-1}\,cm^{-2}\,\text{\AA}^{-1}}$.

\subsection{Invalid Spectrum Filtering}
To remove spectra that contain no meaningful information or lack typical quasar spectral features, we applied the following criteria:
\begin{enumerate}
  \item \textbf{Flat Spectra:} If the flux vector $\boldsymbol{f} \in \mathbb{R}^{4618}$ exhibits no variation across the entire wavelength range:
  \begin{equation}
      \mathrm{ptp}(\boldsymbol{f}) = \max(\boldsymbol{f}) - \min(\boldsymbol{f}) = 0,
  \end{equation}
  the spectrum is considered static and physically uninformative, and is thus discarded.
  
  \item \textbf{Quasi-linear Spectra:} After resampling, if the spectrum can be approximated by a first-order polynomial $f(i) = ai + b$ such that the maximum residual across all wavelengths satisfies:
    \begin{equation}
        \max_i \left| f_i - (ai + b) \right| < \varepsilon,
    \end{equation}
  it is deemed to lack spectral line structures and is removed accordingly. It is noted that $\varepsilon = 10^{-5}$ that we take is a very small threshold, ensuring that spectra with even weak or narrow emission-line features are retained.
\end{enumerate}

\subsection{Spectral Normalization}
To eliminate flux amplitude differences among different targets, all retained spectra were normalized to the $[0,1]$ range using min-max scaling. Given the interpolated flux vector $\hat{\boldsymbol{f}} \in \mathbb{R}^{4618}$, the normalization is defined as:

\begin{equation}
    \tilde{\boldsymbol{f}} = \frac{\hat{\boldsymbol{f}} - \min(\hat{\boldsymbol{f}})}{\max(\hat{\boldsymbol{f}}) - \min(\hat{\boldsymbol{f}})}.
\end{equation}

To further improve model training stability, we also applied zero-mean normalization and unit-norm constraint . The final processed spectra are denoted as \texttt{Norm\_Flux}.

\begin{table*}[htbp]
\centering
\caption{Description of dataset fields}
\label{tab:dataset_fields}
\begin{tabular}{lp{11.5cm}}
\toprule
\textbf{Field} & \textbf{Description} \\
\midrule
\texttt{Defined\_Mass} & Adopted BH mass ($\log (M_{\odot})$), taking the first positive value from \texttt{HA\_MRM}, \texttt{HB\_MRM}, \texttt{MG2\_MRM}, \texttt{C4\_MRM}; set to $-1$ if none. \\
\texttt{Flux} & Raw observed flux on non-uniform grid ($10^{-17}\,\mathrm{erg\,s^{-1}\,cm^{-2}\,\text{\AA}^{-1}}$). \\
\texttt{Flux1} & Flux interpolated onto uniform \texttt{Wave1} grid (same units as \texttt{Flux}). \\ 
\texttt{HA\_MRM} & RM-derived BH mass using H$\alpha$ line lag ($\log (M_{\odot})$). \\
\texttt{HB\_MRM} & RM-derived BH mass using H$\beta$ line lag ($\log (M_{\odot})$). \\
    \texttt{MG2\_MRM} & RM-derived BH mass using Mg\,\textsc{ii} line lag ($\log (M_{\odot})$). \\
\texttt{C4\_MRM} & RM-derived BH mass using C\,\textsc{iv} line lag ($\log (M_{\odot})$). \\
\texttt{Mean\_MJD} & Mean Modified Julian Date of spectroscopic visits. \\
\texttt{Norm\_Flux} & Min-max normalized \texttt{Flux1} (unitless). \\
\texttt{RMID} & Unique ID in SDSS-RM catalog (0-999). \\
\texttt{EPOCH} & Index of spectroscopic observation epoch. \\
\texttt{Rest\_Flux} & Rest-frame flux: \texttt{Flux}/(1+$z$). \\
\texttt{Rest\_Flux\_Normalized} & Min-max normalized rest-frame flux. \\
\texttt{Rest\_Wave} & Rest-frame wavelength: $\lambda_{\rm obs}/(1+z)$ (\AA). \\
\texttt{Wave} & Observed vacuum wavelength (\AA). \\
\texttt{Wave1} & Uniform interpolated wavelength grid (model input). \\
\texttt{ZFINAL} & Adopted redshift of the object. \\
\texttt{Name} & SDSS-DR12 designation (J2000, hhmmss.ss+ddmmss.s). \\
\bottomrule
\end{tabular}
\end{table*}

\subsection{Black Hole Mass Label Construction}
To facilitate supervised learning applications, we adopt reverberation mapped black hole mass estimates from the SDSS-RM spectral analysis pipeline. These RM-based masses, obtained from the emission lines H$\alpha$, H$\beta$, Mg \textsc{ii}, and C\textsc{iv}, are represented as \texttt{HA\_MRM}, \texttt{HB\_MRM}, \texttt{MG2\_MRM}, and \texttt{C4\_MRM}, respectively. For each sample, a unified mass label, \texttt{Defined\_Mass}, was determined by selecting the first valid estimate according to the priority sequence: \texttt{C4\_MRM}, \texttt{MG2\_MRM}, \texttt{HB\_MRM}, and \texttt{HA\_MRM}. If all lines mass estimates are invalid, no defined mass is assigned, and such quasars are excluded from supervised training but remain available for unsupervised spectral reconstruction. These labels are subsequently linked to their corresponding spectra using RMID identifiers and integrated into the dataset.

\subsection{Dataset Partitioning}

To promote robust generalization and reliable performance evaluation, we split the dataset by RMID into training (70\%), validation (15\%), and test (15\%) subsets. The reference set, containing quasars with reverberation-mapped black hole masses, was included only in the training and validation subsets and excluded from the test set to avoid information leakage. 

\section{Methodology}
\label{sec:model}

\subsection{Neural Network}
\begin{figure*}[htbp]
\centering
\includegraphics[width=1\linewidth]{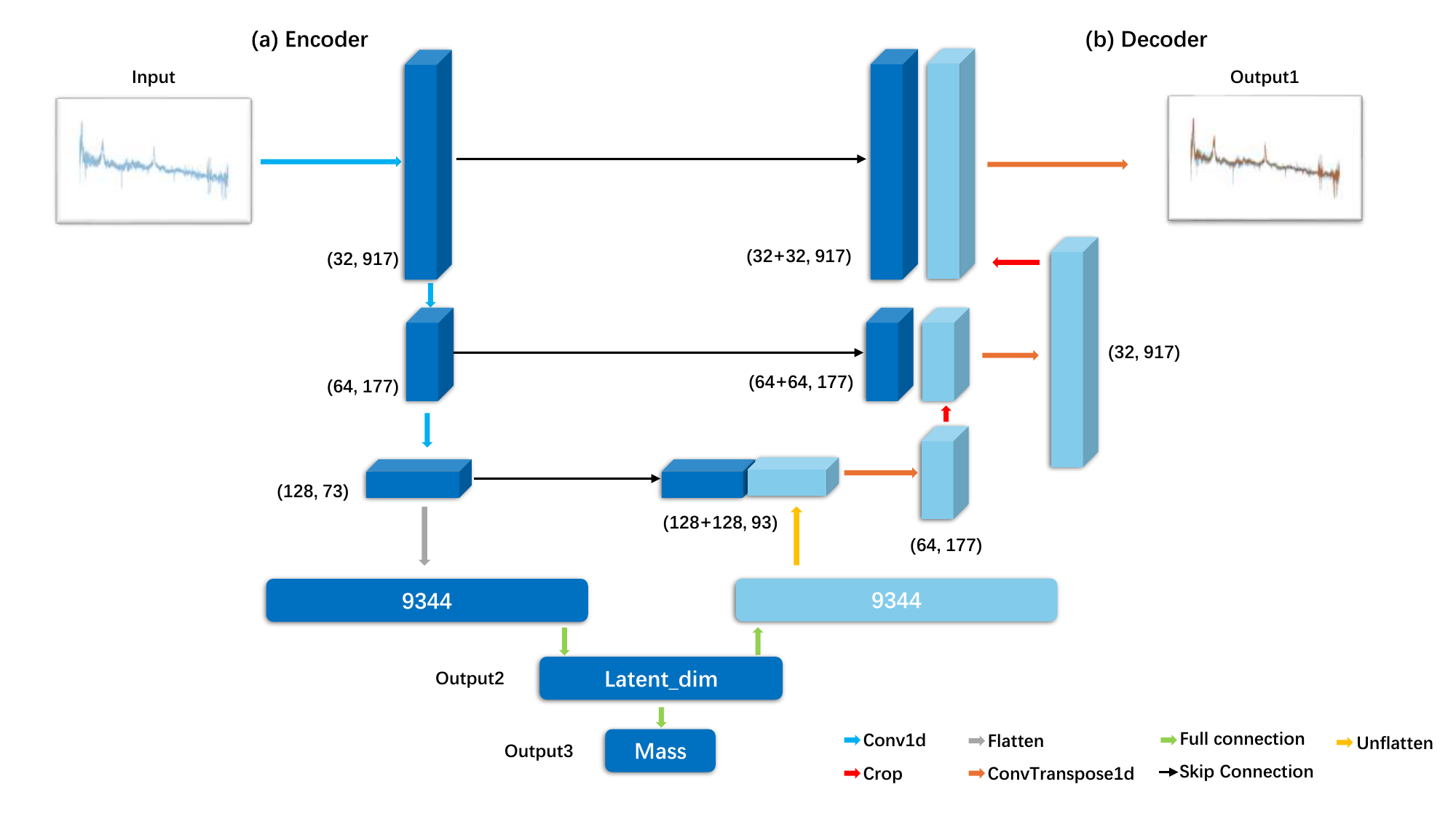}
\caption{Architecture of the proposed deep autoencoder-based neural network.}
\label{fig:model}
\end{figure*}

In this study, we propose a deep autoencoder-based neural network that concurrently performs feature extraction, spectral reconstruction, and black hole mass prediction. The proposed architecture comprises three interconnected modules: an encoder, a decoder, and a dedicated mass prediction branch, as illustrated in Figure~\ref{fig:model}.

The encoder compresses input spectral data into a compact latent representation that encapsulates crucial spectral characteristics. Specifically, the input spectral data, structured as single-channel one-dimensional tensors of dimensions (\textit{batch\_size}, 1, \textit{sequence\_length}), undergo processing by three convolutional blocks. Each block comprises convolutional layers that progressively increase the number of feature channels, followed by batch normalization (BatchNorm1d) for training stabilization, a Rectified Linear Unit (ReLU) activation function for nonlinear transformation, and a dropout layer to mitigate overfitting and enable uncertainty estimation. Subsequently, the resulting feature maps are flattened and projected into a lower-dimensional latent vector of dimension \textit{latent\_dim} via a fully connected layer (\texttt{encoder\_fc}). Additionally, the encoder incorporates a specialized branch for directly predicting black hole mass from these latent representations. This prediction branch involves a fully connected layer that reduces dimensionality to 64, followed by a ReLU activation, a dropout layer (with a dropout probability of 0.5), and finally, a fully connected layer outputting a scalar corresponding to the predicted black hole mass.

The decoder reconstructs the original spectral input from the latent features produced by the encoder. Initially, a fully connected layer expands the latent vector to match the encoder's output dimensions (128$\times$73). The expanded vector is subsequently reshaped by an \texttt{Unflatten} operation into the dimensions (\textit{batch\_size}, 128, 73) suitable for convolutional processing. Reconstruction then proceeds through three transpose convolutional layers, progressively reducing the number of channels and increasing spatial resolution. Critically, these transpose convolutional layers utilize skip connections by concatenating their outputs with corresponding encoder feature maps, preserving detailed spatial information, facilitating gradient flow, and preventing vanishing gradients. The transpose convolutions sequentially reduce channels from 128 to 64, 64 to 32, and finally 32 to 1, with kernel sizes, strides, and padding chosen to ensure output sequence length matches the original spectral data. After concatenation, cropping operations (\texttt{min\_length}) maintain consistent feature map dimensions throughout reconstruction.

\subsection{Physically Motivated Architecture}

We construct the network from the physics of quasar spectra rather than from a purely accuracy-driven objective. Single epoch mass estimators follow the virial form $M_{\rm BH}\propto V^{2}R(L)$, where $V$ is a velocity scale of the broad-line region and $R(L)$ is a luminosity dependent size. This relation is a useful first-order guide, but its accuracy is limited when $R(L)$ is forced to be a simple power law and when line-dependent effects such as profile asymmetry, outflows, and orientation are ignored. Our design goal is to encode the information that enters $V^{2}R(L)$ and, at the same time, allow the model to use additional spectral indicators.

The architecture in Fig. \ref{fig:model} is an encoder--decoder with skip connections. The encoder compresses each spectrum into a low dimensional latent vector intended to carry global, slowly varying information. The decoder reconstructs the spectrum using both the latent representation and intermediate encoder features passed through the skips. The mass regression branch reads only the latent vector, ensuring that the predicted $M_{\rm BH}$ depends on physically interpretable global features rather than on local variations.

Input spectra are min-max scaled and standardized to zero mean and unit variance. Normalization removes the absolute flux scale, so the network cannot recover luminosity from amplitude alone. It must infer luminosity from dimensionless features such as continuum shape, line ratios, and equivalent widths, which are known to correlate with ionizing power (for example, through the radius--luminosity relation and Baldwin-type trends). 
This choice is deliberate: it encourages the encoder to represent physically invariant relations between continuum and lines rather than memorize survey-specific flux levels.

We limit the latent space to be as small as practical to preserve interpretability and control. Quasar spectra are set by a few effective degrees of freedom: a luminosity or radius scale, a velocity scale, the Eddington ratio, broad spectral slope or curvature, and metal-line strength or ionization state. A compact latent vector makes these factors explicit and limits the tendency to spread information across many entangled components. With skip connections, the reconstruction loss no longer forces the latent vector to carry high frequency line details, because the shortest gradient path for reconstruction runs through shallow layers and the skips. The latent representation is therefore shaped mainly by the mass loss and concentrates on slowly varying features. Since the mass head reads only the latent vector, it naturally learns a combination of a luminosity or radius scale with a coarse velocity summary rather than a catalogue of profile minutiae. This aligns with the virial structure of the problem, in which $R(L)$ is a global quantity and $V$ can be summarized by low-order width measures, and it keeps the inferred $M_{\rm BH}$ tied to a physically interpretable global representation rather than to local variability driven by variabilities.

\subsection{Model Training and Optimization}

We adopt a joint optimization approach, simultaneously minimizing two mean squared error (MSE) loss functions: one for spectral reconstruction and the other for black hole mass estimation. These loss functions are combined with equal weights (1:1) and optimized using the Adam algorithm with an initial learning rate of $1\times10^{-4}$ and a batch size of 64. The maximum number of training epochs is set to 500. At each epoch, model performance is evaluated on both the training and validation sets. An early stopping mechanism, with a patience of 50 epochs based on the validation loss, is applied to mitigate overfitting, and the model weights corresponding to the best validation performance are retained.

\section{Evaluation}
\label{sec:evaluation}

\subsection{Effectiveness of neural network for feature extraction}

\begin{figure*}[htbp]
\centering
\includegraphics[width=1\linewidth]{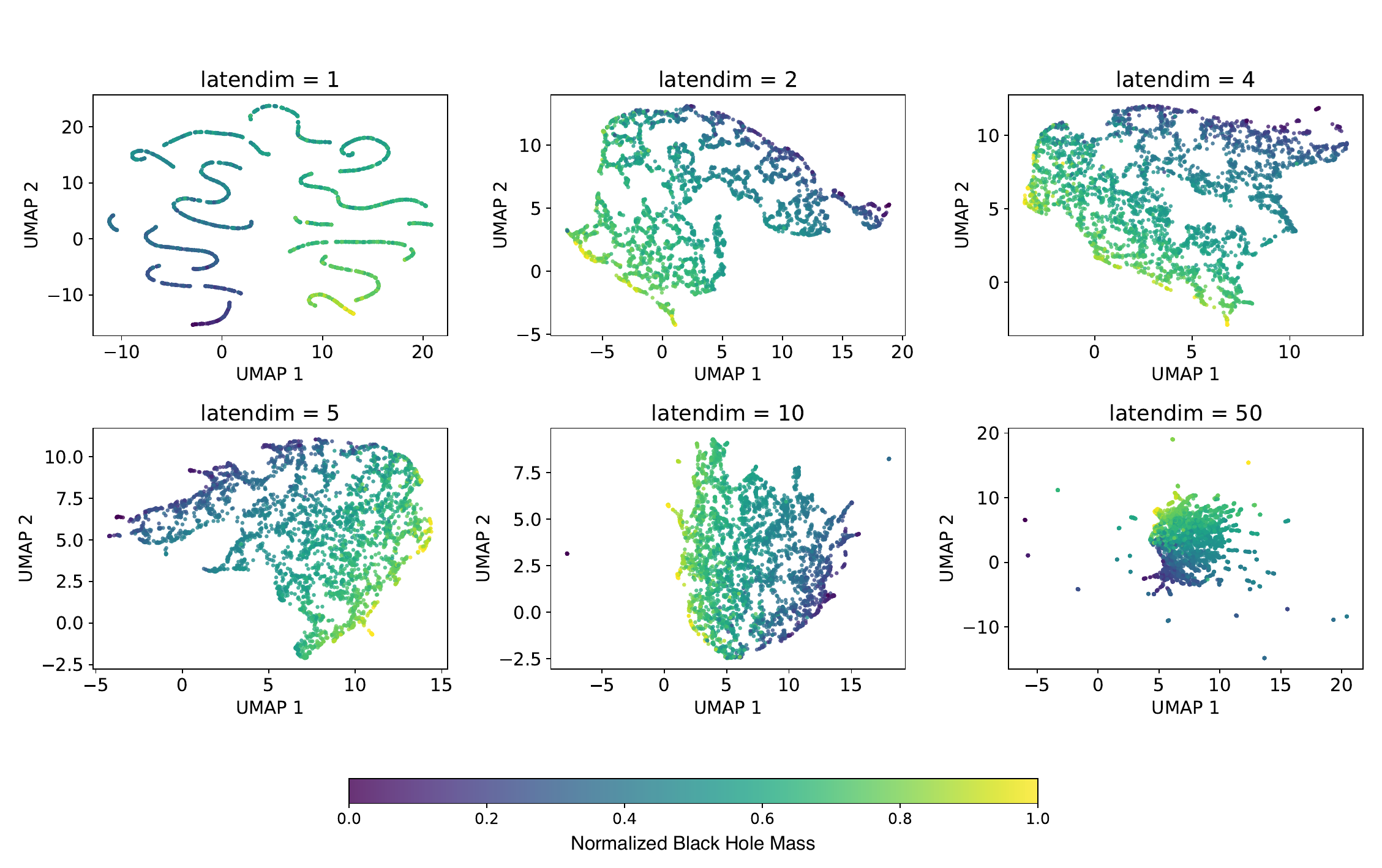}
\caption{
UMAP projections of the learned latent representation for latent dimensionality $d=1,2,4,5,10,$ and $50$. Points are colored by $\log M_{\rm BH}$. A smooth mass gradient is visible in all panels. For $d=1$, the embedding collapses to a single arc with a monotonic gradient; for $d=2$, a principal ridge with transverse structure appears; for $d=4$-$5$, the manifold thickens while the gradient remains continuous; for $d=10$, the global pattern and gradient are preserved with a slightly more compact cloud; for $d=50$, the central manifold persists and a small fraction of points is displaced to the periphery. UMAP preserves local neighborhoods rather than global geometry, so axis orientation is not physically meaningful.}
\label{fig:umap}
\end{figure*}

To evaluate the robustness and interpretability of the learned spectral representations, we trained the autoencoder using different latent space dimensionalities. Varying the latent dimension provides a way to quantify how much spectral-mass information can be retained with minimal model complexity. A lower-dimensional latent space favors physical interpretability, since each latent variable tends to encode broad and physically meaningful spectral variations. Increasing the dimensionality can capture finer spectral details but may also fit stochastic noise and obscure clear physical meaning. Therefore, we aim to achieve an accurate black hole mass estimation with the smallest possible number of latent variables, balancing predictive performance and interpretability. Technically, we employed Uniform Manifold Approximation and Projection (UMAP), a dimensionality reduction technique that maps high-dimensional data to a two-dimensional plane \citep{2018arXiv180203426M}. Specifically, we applied UMAP to project latent representations of dimensions 1, 2, 4, 5, 10, and 50, and visualized the resulting projections by coloring data points according to their normalized logarithmic virial masses ($\log M_{\rm vir}$), as presented in Figure~\ref{fig:umap}.

The results shown in Fig. \ref{fig:umap} visualize how the learned latent representation organizes spectra as the latent dimensionality varies. Points are colored by $\log M_{\rm BH}$, and a smooth ordering with $\log M_{\rm BH}$ is visible at all settings. 

At one latent dimension, the embedding collapses to a single arc with a monotonic $\log M_{\rm BH}$ gradient, showing that the dominant variation relevant to mass is already captured. At two dimensions, the map retains a principal ridge with a clear mass gradient and exhibits transverse structure, indicating that additional modes of spectral variation are represented in the 2D projection. At four and five dimensions, the manifold appears thicker in the UMAP plane, the mass ordering remains continuous along the main ridge. This suggests that added capacity captures secondary structure, for example broad continuum shape or cross line correlations, without disrupting the primary ordering. When the latent dimensionality increases to ten, the global pattern and mass gradient are preserved and the cloud becomes slightly more compact in the UMAP plane, consistent with extra capacity absorbing secondary variance rather than changing the core geometry. At fifty dimensions, the central structure and mass ordering persist, but a small fraction of points is displaced toward the periphery as scattered outliers, which likely correspond to unusual line shapes or lower signal to noise. The scattering reflects overcapacity that allows peripheral directions in latent space to be populated, not a fragmentation of the main manifold. The above analysis supports that five latent representations provide a practical balance between representation power and interpretability.

For evaluate the model's fitting performance, we adopt two metrics: the root mean squared error (RMSE) and the Pearson correlation coefficient $r$.  The RMSE is given by
\begin{equation}
    \mathrm{RMSE}=\sqrt{\frac{1}{N}\sum_{i=1}^N\bigl(\hat y_i - y_i\bigr)^2}\,,
\end{equation}
where $\hat y_i$ and $y_i$ are the predicted and true values for sample $i$, respectively, and $N$ is the total number of samples.  A lower RMSE indicates better predictive performance.

    The correlation coefficient $r$ measures the strength of the linear relationship between predictions and ground truth values:
\begin{equation}
    r=\frac{\sum_{i=1}^N(\hat y_i - \bar{\hat y})\,(y_i - \bar y)}
{\sqrt{\sum_{i=1}^N(\hat y_i - \bar{\hat y})^2}\,
 \sqrt{\sum_{i=1}^N(y_i - \bar y)^2}
}\,,
\end{equation}
where $\bar{\hat y}$ and $\bar y$ denote the means of the predicted and true values, respectively.  A value of $r$ closer to 1 indicates a stronger linear correlation.

The fitting performance results, as presented in Table~\ref{tab:feature_performance}, are characterized by an $R^2_{\text{train}}$ of 0.9570 and an $R^2_{\text{val}}$ of 0.9099. These values indicate strong predictive accuracy on the training data and robust generalization to unseen data. Regarding error metrics, the RMSE is 0.00310 for the training set and 0.05532 for the validation set, with an MAE of 0.03896 on the validation set. These metrics, which are among the most favorable in Table~\ref{tab:feature_performance}, demonstrate that the model achieves high accuracy in predicting black hole masses with the chosen five-dimensional latent space.

Therefore, considering both interpretability and predictive performance, we adopt five latent dimensions as the optimal configuration. The five-dimensional representation preserves a coherent mass gradient while effectively suppressing noise and non-physical spectral variations. Quantitative regression tests further support this choice, showing that increasing the dimensionality beyond 5 does not bring significant improvement in the mass fitting accuracy.

The physical analysis and the cosmological studies based on these results will be the main subjects of the subsequent papers.


\begin{table*}[htbp]
\centering
\caption{Model performance for different numbers of latent features. The table lists the coefficient of determination ($R^{2}$), root-mean-square error (RMSE), and mean absolute error (MAE) of training and validation datasets for different latent space dimensionalities. The results show that the predictive accuracy improves rapidly up to $\sim$5 latent dimensions and then grows slowly, indicating that five dimensions provide the optimal balance between model complexity and performance.}
\label{tab:feature_performance}
\begin{tabular}{lccccc}
\toprule
Latent Dim &
  $R^2_{\rm train}$ &
  $R^2_{\rm val}$ &
  RMSE$_{\rm train}$ &
  RMSE$_{\rm val}$ &
  MAE$_{\rm val}$ \\
\midrule
1  & 0.9265 & 0.8549 & 0.00512 & 0.07154 & 0.05329 \\ 
2  & 0.9432 & 0.9075 & 0.00326 & 0.05712 & 0.03920 \\ 
4  & 0.9495 & 0.9081 & 0.00324 & 0.05693 & 0.03911 \\ 
5  & 0.9570 & 0.9099 & 0.00310 & 0.05532 & 0.03896 \\
10 & 0.9579 & 0.9132 & 0.00306 & 0.05534 & 0.03768 \\ 
50 & 0.9568 & 0.9142 & 0.00303 & 0.05503 & 0.03617 \\ 
\bottomrule
\end{tabular}
\end{table*}

\begin{figure*}[htbp]
  \centering
  \includegraphics[width=0.8\linewidth]{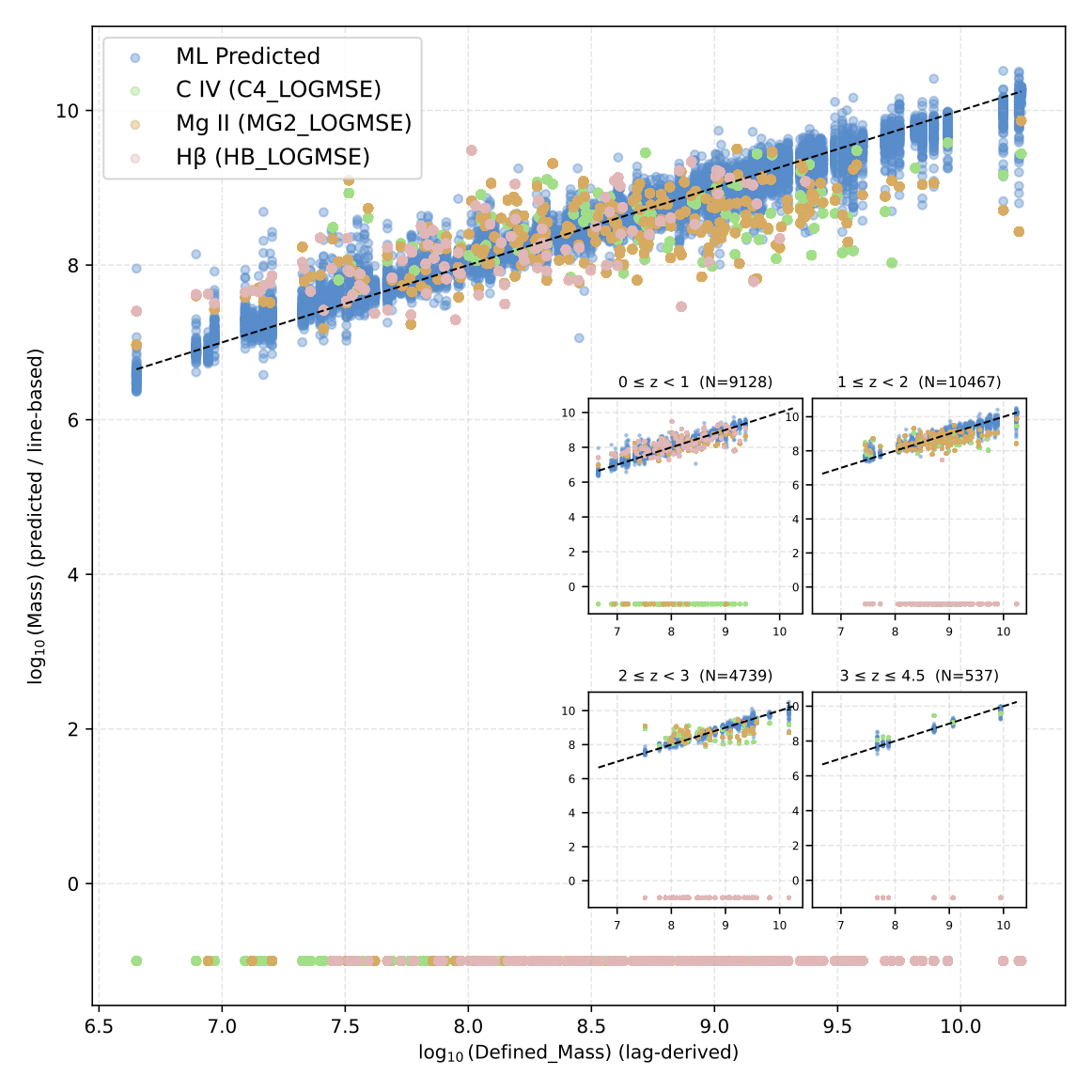}
  \caption{Comparison between black hole mass estimates from our autoencoder-based model and those obtained using traditional single-epoch virial methods for the test dataset. The blue points show the predictions from the neural-network model, while the purple, yellow, and green points correspond to single-line estimates based on H$\beta$, Mg\,\textsc{ii}, and C\,\textsc{iv}, respectively. The dashed line marks the one-to-one relation expected for perfect agreement with the reverberation-mapping (RM) masses. The scatter of the blue points mainly reflects variations among individual spectra from the multi-epoch SDSS-RM observations, as each spectrum is treated independently in the estimation. Traditional estimators exhibit larger scatter and systematic deviations, particularly at the low- and high-mass ends, and many sources lack reliable single-line measurements. In contrast, our model maintains a tight correlation with the RM masses ($R^{2}=0.909$) and achieves a low overall RMSE of 0.058\,dex with respect to the RM-based mass, and also providing predictions for objects where the conventional methods fail.}
  \label{fig:Predicted}
\end{figure*}


\begin{figure*}[htbp]
  \centering
  \includegraphics[height=10cm]{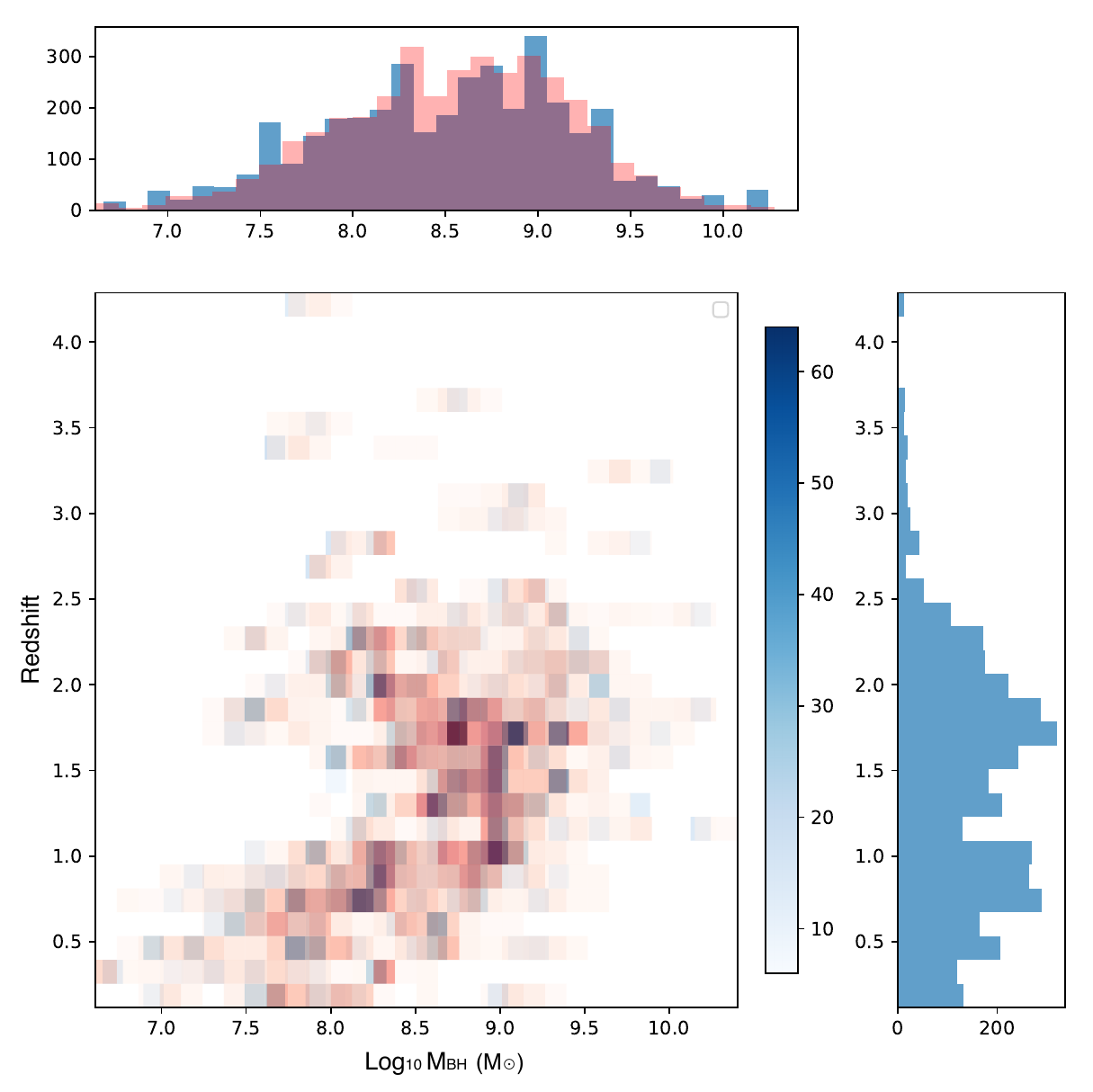}
  \caption{Heatmap of predicted versus RM-based black hole masses across different redshifts, with two overlaid layers:  The blue colormap shows RM-based masses; the semi-transparent red colormap shows network-predicted masses.}
  \label{fig:distribution}
\end{figure*}
\subsection{Redshift-mass coverage and consistency with SDSS-RM} \label{subsec:rm-coverage} 
Figure~\ref{fig:distribution} visualizes our predictions on the SDSS-RM sample as a mass-redshift heatmap overlaid on the RM-based masses. The two layers show close agreement across the full dynamic range, with both distributions peaking near $z\!\sim\!1$-2 and spanning $\log_{10}(M_{\rm BH}/{\rm M_\odot})\!\sim\!7$-10. This pattern is consistent with the coverage and mass range reported in the SDSS-RM Key Results release, which summarizes the RM sample over $0\!\lesssim\!z\!\lesssim\!4$ and typical black-hole masses of $10^{7}$-$10^{9.5}\,{\rm M_\odot}$ \citep{2011ApJS..194...45S,2020ApJS..250....8L,shen2024sloan}. The agreement indicates that our network reproduces the established population-level trends in the SDSS-RM dataset without introducing apparent selection-driven distortions. 

\subsection{Comparison with Single-Line Virial Estimators}

The model's predictions on the validation set are highly consistent with the RM based black hole masses, with the scatter plot (figure \ref{fig:Predicted}) tightly clustered around the ideal consistency line. In other words, the model does not introduce significant systematic biases and shows no evidence of overfitting-related biases, which strengthens our confidence in its generalizability. 


\begin{table*}[htbp]
\centering
\small 
\setlength{\tabcolsep}{4pt} 
\caption{Summary of Black Hole Mass Prediction Methods. The table lists the RMSE (in dex) and $R^{2}$ evaluated against RM-based masses as the reference, reported in the respective studies.}
\label{tab:different}
\begin{tabular}{l r r l} 
\toprule
\textbf{Method} & \textbf{RMSE (dex)} & \textbf{$R^2$} & \textbf{Data} \\
\midrule
Single Epoch Virial\citep{Vestergaard} & 0.35 & -- & Single epoch spectra \\
AGNet Deep Learning\citep{2023MNRAS.518.4921L} & 0.371 & 0.428 & Photometric time series \\
X-ray Reverberation NN\citep{2022MNRAS.513..648C} & 0.1077 & 0.9124 & X-ray light curves \\
Lasso Regression\citep{2022RAA....22h5014H} & 0.50 & -- & Photometry + morphology \\
CQR\citep{2023MNRAS.524.3116Y} & \mbox{0.198 (H$\beta$); 0.222 (Mg\,II)} & \mbox{0.7 (H$\beta$); 0.85 (Mg\,II)} & SDSS single epoch spectra \\
\textbf{QuasarSpecNet (This Work)} & \textbf{0.058} & \textbf{0.909} & \textbf{SDSS spectra} \\
\bottomrule
\end{tabular}
\end{table*}

Table~\ref{tab:different} compares our method with traditional single-line virial estimators based on H$\beta$, Mg\,\textsc{ii}, and C\,\textsc{iv} scaling relations, as well as with other machine learning approaches. Conventional single-epoch relations typically generate scatters of $\sim$0.2--0.5 dex and are prone to systematic biases \citep{2017MNRAS.465.2120C}, particularly for C\,\textsc{iv}-based estimates in high-accretion quasars. In contrast, our model achieves an overall RMSE of only 0.058 dex, representing almost an order-of-magnitude improvement in precision. Our model also achieves a high $R^2 > 0.9$, indicating reduced scatter and improved accuracy across the full sample.

The comparison between our autoencoder-based predictions and the traditional single-line virial estimators is shown in Figure~\ref{fig:Predicted}. Our predictions (blue) closely follow the one-to-one relation across the full dynamic range in mass, demonstrating excellent agreement with the RM masses. In contrast, the single-line estimates (H$\beta$: purple; Mg,\textsc{ii}: yellow; C,\textsc{iv}: green) exhibit larger systematic deviations at both the low- and high-mass ends, which is a known problem in previous studies. At the high-mass end, C,\textsc{iv}-based estimates are strongly affected by non-virial outflows and blueshifts that distort the line width and centroid, leading to overestimation of the black hole mass \citep[e.g.,][]{2013ApJ...775...60D,2017MNRAS.465.2120C}. At the low-mass end, the weaker and noisier broad lines, combined with uncertainties in continuum subtraction and line fitting, increase measurement scatter and bias \citep{2013ApJ...775...60D,2008ApJ...680..169S}. Additional  differences arise from variations in the virial factor and broad-line region geometry across luminosity and accretion states \citep{2017FrASS...4...70M}. Moreover, sample selection and luminosity-dependent Malmquist bias can further amplify these edge effects in flux-limited surveys \citep{2008ApJ...680..169S}. 

A further advantage of our approach is its ability to recover reliable black hole masses even when prominent emission lines are weak, noisy, or fall outside the observed spectral range, where traditional virial estimators fail. For instance, H$\beta$-based estimates are unavailable for high-redshift quasars where the line is redshifted beyond the optical coverage of SDSS, and Mg,\textsc{ii}-based estimates become inaccessible beyond $z \sim 2$ when the line shifts into the near-infrared \citep[e.g.,][]{2008ApJ...680..169S}. In contrast, C,\textsc{iv}-based estimates remain available at higher redshift but are strongly affected by line blueshifts and non-virial outflow components, leading to substantial scatter and systematic bias in the inferred masses \citep[e.g.,][]{2013ApJ...775...60D,2017MNRAS.465.2120C}. These limitations highlight the advantage of our model, which leverages the full spectral information to obtain consistent mass estimates even in regimes where single-line methods are unreliable or inapplicable.


\section{287,872 Black Hole Mass} \label{sec:287,872 Black Hole Mass}

\begin{figure}[]
    \centering
    \includegraphics[width=1\linewidth]{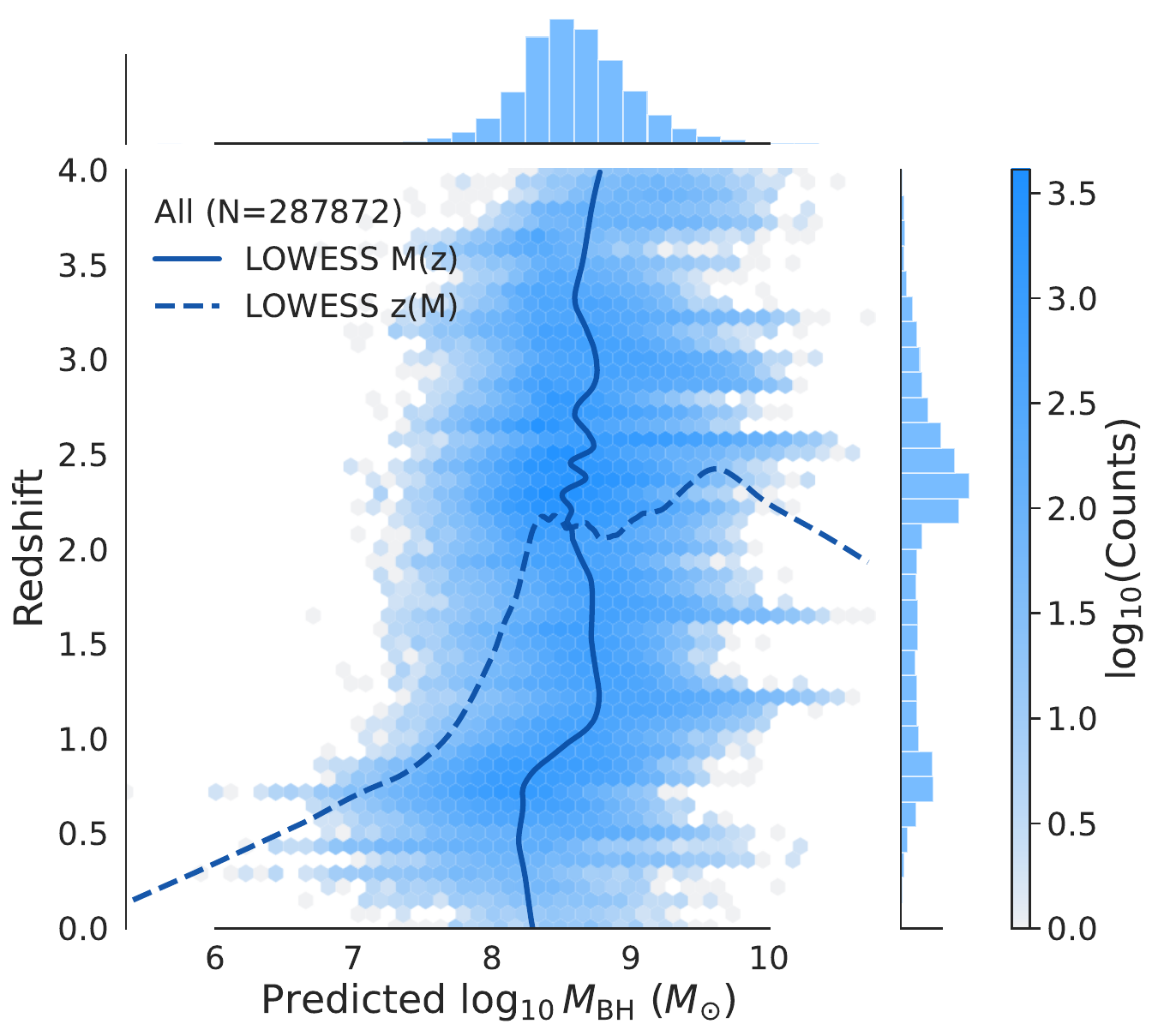}
    \caption{SDSS DR16 quasar sample: hexbin map of predicted black-hole mass versus redshift, with a logarithmic color scale for counts. Two \textsc{LOWESS} curves are overplotted: solid for $\log M$ as a function of $z$, dashed for $z$ as a function of $\log M$, tracing the central trend. Top and right panels show marginal histograms. The redshift is limited to $0 \le z \le 4$.}
    
    \label{fig:DR16_result}
\end{figure}

\begin{figure}
    \centering
    \includegraphics[width=1\linewidth]{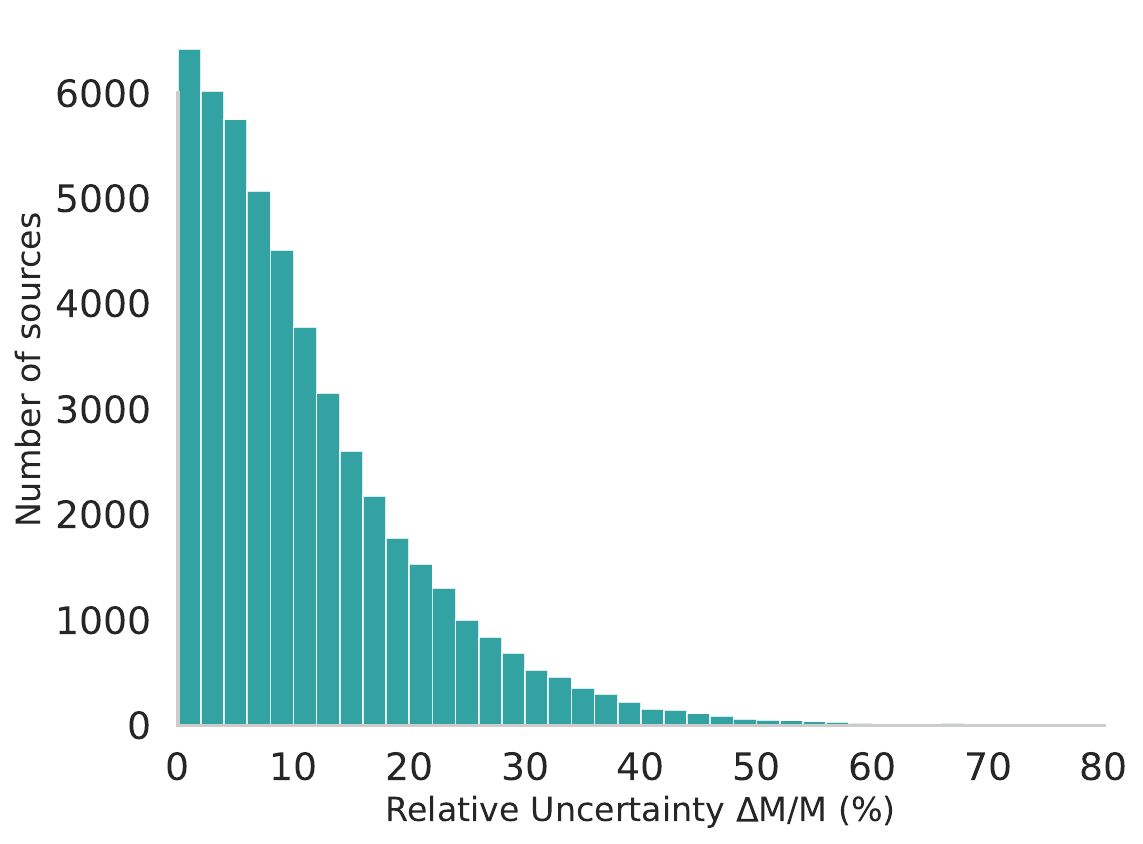}
    \caption{Distribution of the relative standard deviation ($\Delta$M/M) of black hole mass estimates for sources with multiple spectroscopic observations. The majority of sources show a relative uncertainty below $\sim 10\%$, indicating high internal consistency of the trained network.}
    
    \label{fig:relative_uncertainty}
\end{figure}

\begin{figure}
    \centering
    \includegraphics[width=1\linewidth]{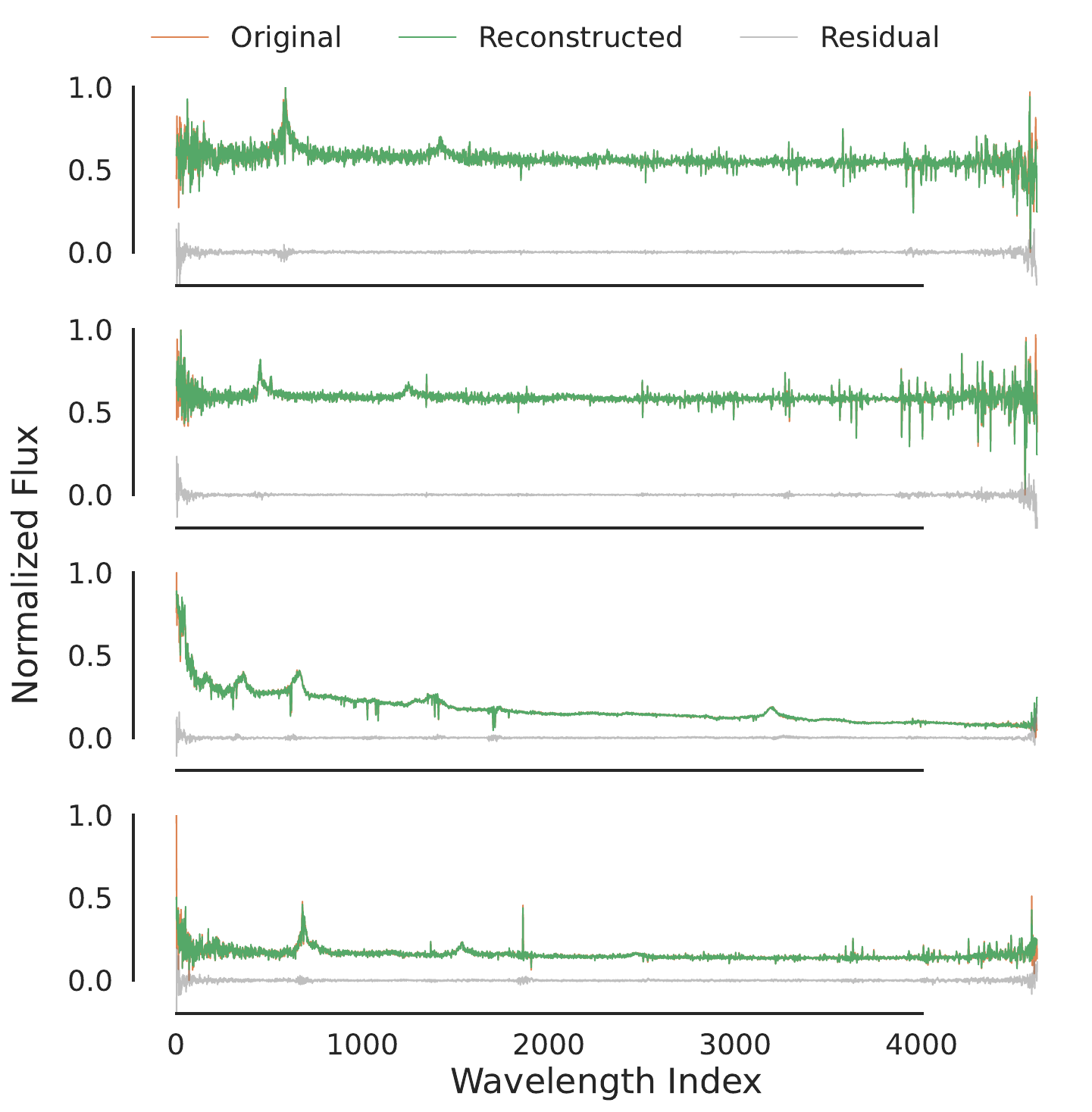}
    \caption{Comparison between original and reconstructed spectra for four randomly selected sources. The reconstructed spectra (green) nearly coincide with the original observed spectra (orange) over the entire wavelength range, showing the network’s high reconstruction accuracy. The residuals (gray) remain small except near the spectral edges, where the instrumental uncertainties dominate.}
    
    \label{fig:example_reconstruction}
\end{figure}

After training, the network can be applied to the large SDSS quasar sample. The SDSS-RM program is a dedicated subproject of SDSS that uses the same spectroscopic instrumentation and reduction pipeline as the main SDSS survey. The difference lies in the observing strategy: while the main SDSS typically provides a single-epoch spectrum for each quasar, SDSS-RM obtained repeated spectra over many epochs, enabling the measurement of emission-line lags and accurate reverberation-mapped black hole masses. Our model is trained on these RM-based labels, but once trained it requires only a single spectrum as input. This allows the network to achieve RM-level accuracy when applied to single-epoch spectra from the main SDSS quasar survey. In this work, we apply the trained model to 287,872 quasars of 377,549 spectra spanning $0 \lesssim z \lesssim 4$, providing a very large catalog of massive black hole mass.

The result is shown in Fig.~\ref{fig:DR16_result}, that the two dimensional distribution of black hole mass as a function of redshift. The inferred black hole masses cover a broad range from $10^{6}\,M_\odot$ to $10^{10}\,M_\odot$. The overall trend agrees with the statistical behavior found in large optical quasar samples \citep[e.g.,][]{2011ApJS..194...45S,2020ApJS..250....8L}, where the most luminous and massive quasars cluster around $z\sim2$--$3$. The persistence of high inferred $M_{\rm BH}$ beyond $z\approx3$ is physically reasonable, since only the most massive and luminous accreting black holes can remain above the survey flux limit at those distances. The absence of a turnover suggests that the neural network estimator does not inherit  the redshift dependent bias that affects virial masses based solely on C\,\textsc{iv}, which often underpredict masses for high blueshift quasars  \citep[e.g.,][]{2017MNRAS.465.2120C,2018MNRAS.478.1929M}. Instead, the model appears to combine continuum and multiband spectral information to maintain a consistent scaling of $M_{\rm BH}$ with redshift.

To evaluate the internal consistency of the model predictions, we analyzed sources with multiple independent spectroscopic observations. A total of 49,404 sources were repeatedly observed. For each source, the standard deviation and relative standard deviation of the predicted black hole masses were calculated, see Fig. \ref{fig:relative_uncertainty}. The model shows a median dispersion of 0.046 dex in black hole mass estimates, corresponding to a median relative uncertainty of 8.6\%. This small internal scatter indicates that the model provides stable and consistent mass estimates across repeated observations, supporting the reliability of its predictions.

The four randomly selected spectra shown in Fig. \ref{fig:example_reconstruction} compare the original SDSS spectra with those reconstructed by the neural network. The reconstructed spectra almost completely overlap with the originals across most of the wavelength range, indicating that the model accurately reproduces both the continuum shape and emission-line structures. The small and structureless residuals confirm the high reconstruction fidelity of the model. The slightly larger discrepancies seen at the shortest and longest wavelengths are expected from the observational characteristics of the SDSS spectrographs. At the spectral edges, the instrumental response and CCD quantum efficiency decrease, the flux calibration becomes less reliable, and the SNR ratio drops sharply. These instrumental effects dominate the uncertainty in those regions, rather than the limitation of the neural network itself.

\section{Discussion}
\label{sec:discussion}

\subsection{Differential Mass function}

\begin{figure}
    \centering
    \includegraphics[width=1\linewidth]{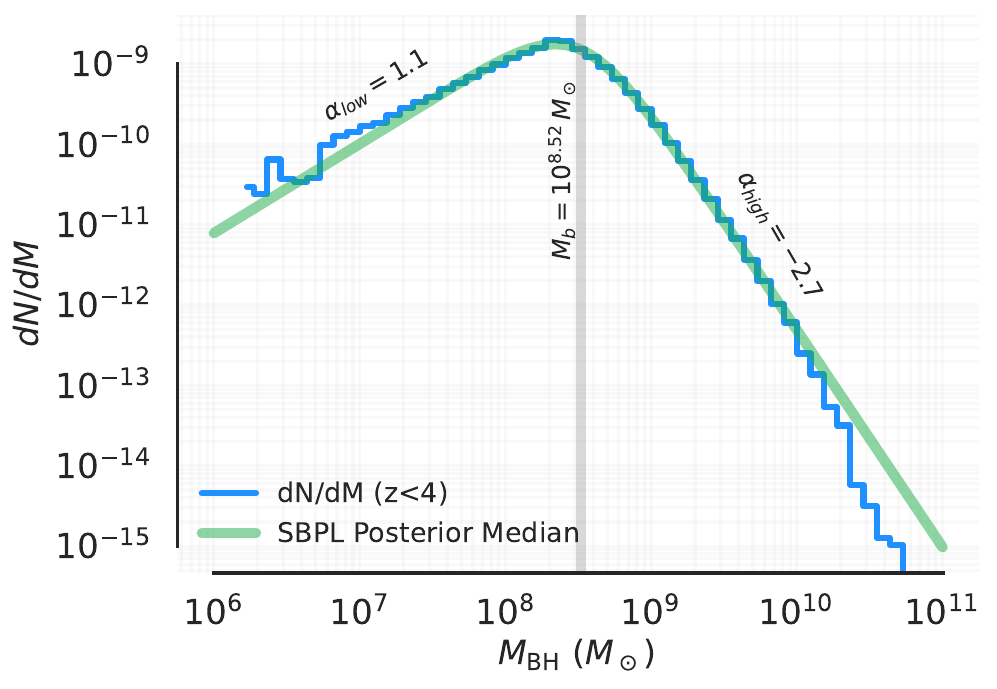}
    \caption{Differential black hole mass function for sources with $z<4$. Blue steps show the binned estimate of $\mathrm{d}N/\mathrm{d}M$; the green curve is the posterior median of SBPL fit obtained with a Poisson likelihood and MCMC sampling. The fit gives a break at $\log_{10}(M_b/M_\odot)=8.52$ (vertical gray line), a low–mass logarithmic slope $\alpha_{\rm low}=+1.1$ and a high–mass slope $\alpha_{\rm high}=-2.7$. }
    
    \label{fig:mass-function}
\end{figure}

We model the differential black hole mass function, $ \mathrm{d}N/\mathrm{d}M $, see Fig. \ref{fig:mass-function}, for all sources with $ z < 4 $, using a smoothed broken power-law (SBPL) of the form  

\begin{equation}
\frac{\mathrm{d}N}{\mathrm{d}M} = A 
\left[
\left( \frac{M}{M_b} \right)^{p_1 s}
+
\left( \frac{M}{M_b} \right)^{p_2 s}
\right]^{1/s},
\end{equation}
where $ A $ is the normalization, $ M_b $ is the break mass, $ p_1 $ and $ p_2 $ are the slopes below and above the break, and $ s $ controls the smoothness of the transition.  The fitting is performed using a Poisson-binned maximum-likelihood estimator followed by an MCMC analysis to sample the posterior parameter distributions. While the total black hole mass function in inactive galaxies follows a Schechter-like form with an exponential cutoff \citep{1976ApJ...203..297S}, the active population is shaped by the convolution of this total distribution with a mass-dependent duty cycle and a broad Eddington-ratio distribution. This convolution  transforms the exponential cutoff into a smoother turnover, producing a double power-law form \citep{2006ApJ...648..128K, 2007ApJ...667..131G, 2010A&A...516A..87S}.

The best-fit model gives a normalization of $ \log_{10}A = -2.84 $, a low-mass slope of $ p_1 = 2.70 $, a high-mass slope of $ p_2 = -1.11 $, a break mass of $ \log_{10}(M_b/M_\odot) = 8.52 $, and a smoothness parameter $ s = 0.58 $.  In logarithmic space, the corresponding differential slopes are $ \alpha_{\mathrm{low}} = +1.1 $ and $ \alpha_{\mathrm{high}} = -2.7 $.  Indicating roughly $\sim10\%$ of the sample has mass more than $10^{9}\,M_\odot$, and only $\sim 0.1\%$ exceeds $10^{10}\,M_\odot$.

The SBPL fit reveals a clear turnover at $\log_{10}(M_b/M_\odot) = 8.5$ ($ M_b = 3\times10^8\,M_\odot $).   Below this mass, the rising slope  reflects the increasing number of moderate-mass black holes, consistent with the expected population of typical quasars.  
At higher masses, the steep decline reflects the relative scarcity of the most massive quasars. A mild excess of sources below $ M_{\mathrm{BH}} \lesssim 10^{7.5}\,M_\odot $ likely originates from low-redshift, low-luminosity quasars that dominate flux-limited samples.  
Conversely, the observed deficit at $ M_{\mathrm{BH}} \gtrsim 10^{10}\,M_\odot $ is plausibly primarily due to a feedback-limited upper bound on black hole growth set by gas supply and radiation pressure, or/and selection incompleteness for extremely luminous objects near the bright end of the quasar luminosity function.

\subsection{Broad and Narrow Line Quasars}

RM requires measurable time delays between the variable continuum and the response of the broad emission lines that originate from the broad-line region (BLR). The BLR gas lies within light-days to light-weeks of the central ionizing source and exhibits Doppler-broadened velocities of several thousand km,s$^{-1}$, producing the broad line profiles seen in Type~1 AGNs and quasars. In contrast, the narrow-line region (NLR) is located at much larger radii (tens to thousands of parsecs) and is photoionized by the central source but does not respond measurably to continuum variations on humanly observable timescales. Its line widths are only a few hundred km,s$^{-1}$, reflecting the gravitational potential of the host galaxy rather than that of the black hole. As a result, narrow emission lines show little or no correlated variability with the continuum, making them unsuitable for RM-based black hole mass measurements. Therefore, the RM-based masses, including those in the SDSS-RM project, are derived from broad-line AGNs, for which the BLR variability can be directly observed and modeled \citep[e.g.,][]{1982ApJ...255..419B,1993PASP..105..247P,2015ApJS..216....4S,2011ApJS..194...45S,shen2024sloan}.


In our framework, the RM-based black hole masses serve as the ground-truth labels for supervised training, while the model input consists of single-epoch spectra. Hence, the neural network does not rely on the temporal information required by reverberation mapping and, in principle, is not restricted to sources showing variable broad lines. However, since all RM measurements used for training originate from broad-line AGNs, the learned spectral representations inherently reflect the some physical conditions of the broad-line region. To assess possible extrapolation beyond this regime, we further divide all quasars into broad-line and narrow-line classes based on their emission-line widths. For broad-line quasars, the model performance on the previous validation set should directly reflect its accuracy. For narrow-line quasars, there is no prior benchmark for true RM-equivalent masses, and the model's reliability cannot be guaranteed a priori since the network is almost a black box. Nevertheless, given that the neural network captures, not only the emission lines features, but also the spectral details and global correlations, it is plausible that its predictions for narrow-line quasars remain physically meaningful. Comparing the inferred masses between the broad- and narrow-line populations may therefore provide an empirical estimate of the model's generalization accuracy outside the RM-calibrated domain.


\begin{figure*}[htbp]
    \centering
    \includegraphics[width=0.48\linewidth]{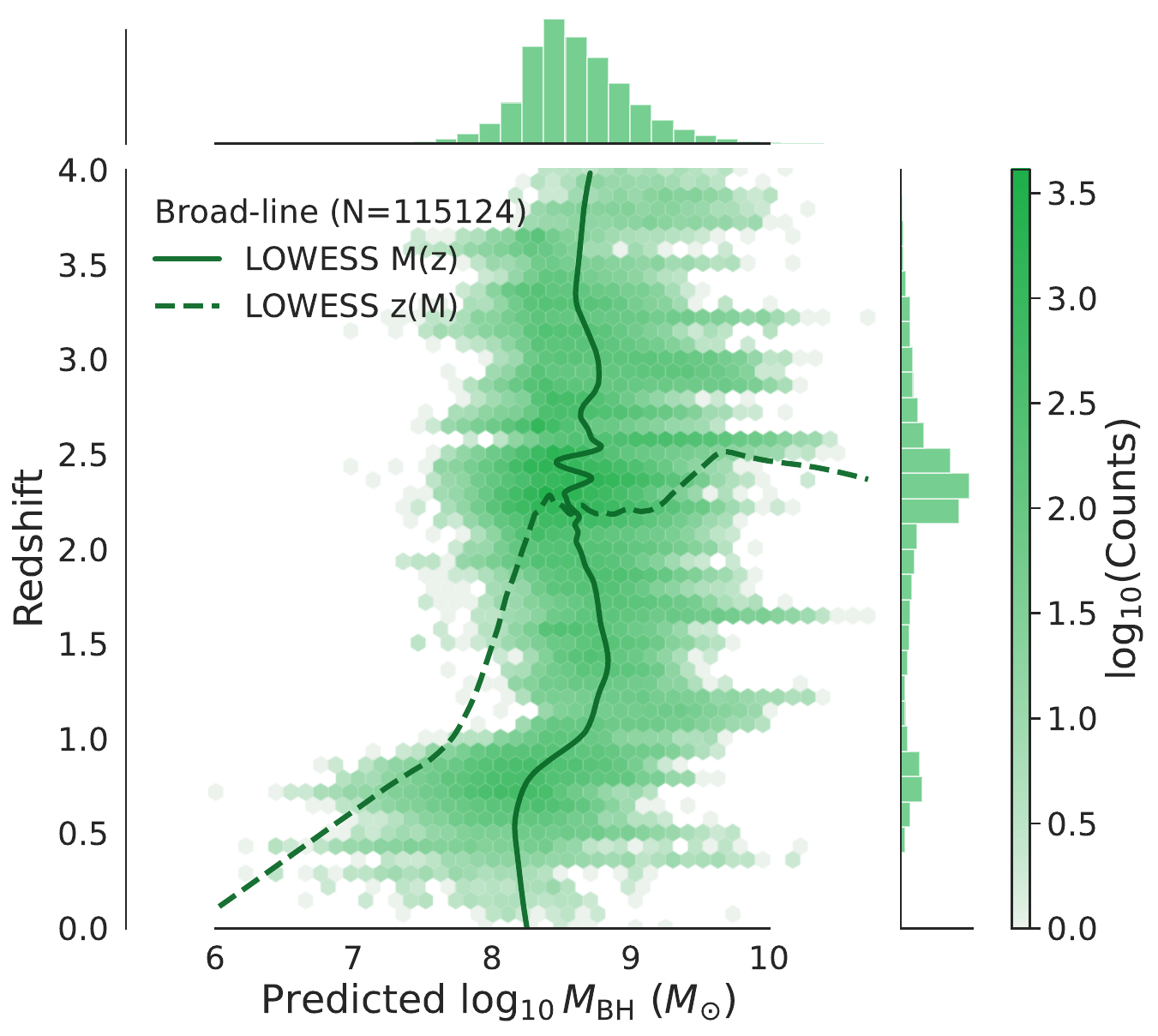}
    \includegraphics[width=0.48\linewidth]{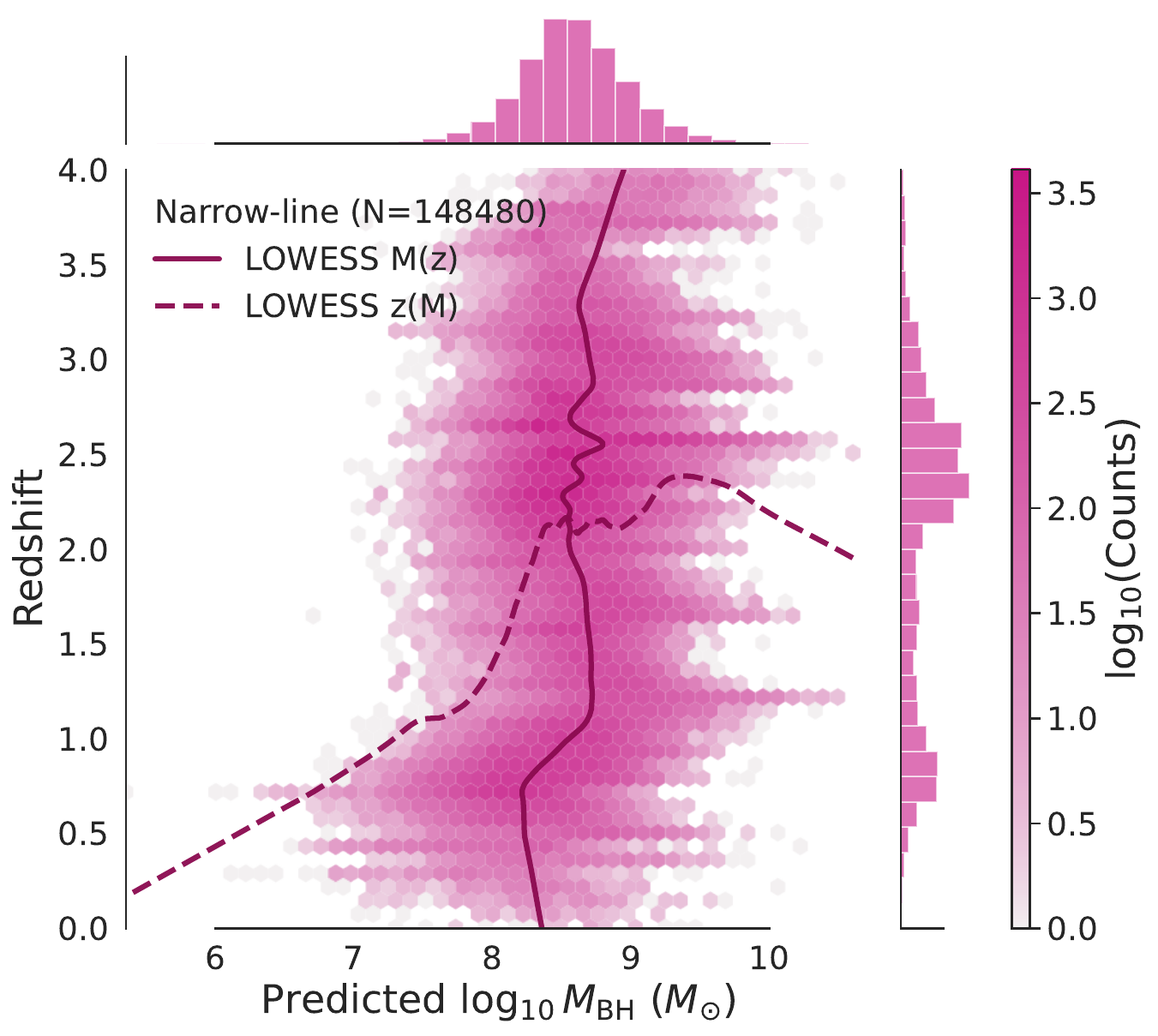}
    \caption{Distribution of broad-line (left) and narrow-line (right) quasars in the redshift-mass plane.Each panel shows a hexbin density map with logarithmic color scaling. Two LOWESS trends are plotted (solid: $\log M$ vs.~$z$; dashed: $z$ vs.~$\log M$), with marginal histograms shown on the top and right. The redshift is limited to $0 \le z \le 4$, and the source number is indicated as N.}
    \label{fig:broad-narrow}
\end{figure*}



Fig.~\ref{fig:broad-narrow} shows the distribution of the predicted black hole mass as a function of redshift for the broad line and narrow line subsamples. Both panels share the same dynamic range in mass and redshift 
($7 \le \log (M_{\rm BH}/M_\odot) \le 10$ and $0 \le z \le 4$). The overall envelope of the two distributions is remarkably similar, indicating that the network assigns consistent mass scales to quasars regardless of their spectral classification. For both subsamples, the source density peaks at $z\simeq2.5$, corresponding to the epoch of maximum quasar activity, and the median mass increases from $\log M_{\rm BH}\approx7.5$ at low redshift to about $\log M_{\rm BH}\approx9.0$ near the peak.

At $z\lesssim3$, the LOWESS trends of the two subsamples nearly coincide, showing a common rise of the typical black hole mass with increasing redshift. Beyond $z\approx3$, a mild divergence appears: the mean mass of the narrow line subsample increases toward high redshift, while the mean of the broad line subsample exhibits a slower increase. This difference is modest compared to the overall scatter and does not represent a discontinuity in the mass distribution. Instead, it likely reflects observational selection effects. The selection of broad line samples requires a SNR>5 and $\mathrm{FWHM} > 1000~\mathrm{km~s^{-1}}$. At $z \gtrsim 3$, the Balmer and Mg\,\textsc{ii} lines move out of the optical band, while C\,\textsc{iv} and the UV continuum dominate the observable region, and the SNR degrades. Very broad lines with shallow wings are preferentially lost from the measurements. This truncates the high-$\mathrm{FWHM}$, high-mass tail in the broad line subsample and pulls down its mean mass. At the same time, C\,\textsc{iv} is affected by outflows that shift the line centroid blueward and distort the profile. These effects can bias the measured $\mathrm{FWHM}$ to smaller values and push some intrinsically Type~1 objects below the broad line threshold, moving them into the narrow line bin and raising its mean mass.

The broad-line subsample also shows a noticeably narrower mass distribution compared to the narrow-line subsample, and it lacks the low-mass tail present in the latter. This asymmetry is likewise a consequence of selection effects. Low-mass black holes tend to have lower luminosities, which reduce the equivalent width and signal-to-noise ratio of their emission lines. As a result, their broad components are more easily missed or fall below the adopted selection thresholds. 

To conclude, the similarity of the two distributions demonstrates that the model produces internally consistent mass estimates for different quasar types.

\subsection{Redshift Distribution}

Fig.~\ref{fig:DR16_result} shows that the overall black hole mass distribution approximately follows a log-normal shape as expected. In contrast, the redshift distribution exhibits several pronounced peaks. The clustering of quasars around $ z \approx 2-2.5 $ arises because strong ultraviolet emission lines such as Mg\,\textsc{ii}~$\lambda2798$ and C\,\textsc{iv}~$\lambda1549$ enter the SDSS optical wavelength range (3800-9200~\AA) in this redshift interval, making quasars in this range easier to identify and classify spectroscopically. The SDSS quasar target selection algorithm, which relies on broadband colors, is also most efficient in this regime since these emission features significantly affect the $ u, g, r, i, z $ colors, increasing selection completeness. The secondary rise at lower redshift ($ z \lesssim 0.8 $) is produced by quasars with strong optical Balmer emission lines, particularly H$\beta$~$\lambda4861$ and H$\alpha$~$\lambda6563$, which fall within the SDSS spectral window and produces high signal-to-noise detections. Together, these effects produce peaks in the observed redshift distribution that reflect both instrumental sensitivity and the redshift dependent efficiency of the quasar selection function.

\subsection{High Redshift Limitations}

The RM dataset contains only 3 quasars at $z>4$ and does not contain any quasars with $z>4.5$ or $\log M_{\rm BH}>10.5$. Since our network is supervised by RM labels, its accuracy depends on the size and coverage of the training sample. With so few high-$z$ sources, it is difficult to guarantee predictive accuracy after training.

In addition, spectra at high redshift differ substantially from those at low redshift. Rest-frame optical lines such as H$\beta$ and Mg\,\textsc{ii} move out of the observed wavelength window, while the far-UV dominates, including lines such as Ly$\alpha$, C\,\textsc{iv}, and Si\,\textsc{iv}+O\,\textsc{iv}]. These features exhibit different systematics, including stronger winds, larger blueshifts, and generally lower signal-to-noise ratios\citep{2002AJ....124....1R,2023MNRAS.526.3967M,2013ApJ...775...60D} This domain shift means that the mapping learned from low-$z$ RM spectra does not transfer cleanly to $z>4$.

Therefore, although the network can produce mass estimates for high-redshift AGN, we do not include these predictions in the main results due to the limited training support  at high $z$.

\section{Conclusion}\label{sec:Conclusion}

We trained a single epoch spectral model using RM masses as labels, then applied it to a uniform sample of 287{,}872 SDSS quasars. This combines RM level calibration with survey scale application, maintaining accuracy while reaching a massive sample size. The predictions track the RM masses with an RMSE scatter of 0.058~dex, which corresponds to a relative uncertainty of about 14\%. From the full catalog we measured the redshift evolution of the black hole mass distribution and derived the differential mass function $\mathrm{d}N/\mathrm{d}M$ for $z<4$. A smoothed broken power-law fit obtains a break at $M_b\simeq3\times10^8\,M_\odot$, a low-mass index $\alpha_{\rm low}\approx+1.1$, and a high-mass index $\alpha_{\rm high}\approx-2.7$. The cumulative form implies that roughly $10\%$ of objects have $M_{\rm BH}>10^{9}\,M_\odot$, while fewer than $10^{-3}\!-\!10^{-4}$ exceed $10^{10}\,M_\odot$.

From the machine-learning perspective, the network is constructed to refine the virial structure $M_{\rm BH}\propto V^{2}R(L)$. An encoder-decoder with skip connections allows high frequency line details to be used for spectral reconstruction, while a compact latent vector carries the global information that is most predictive for mass. We adopt a five-dimensional latent representation to preserve interpretability and to align with physically meaningful factors, for example continuum shape, broad line width statistics, and cross line correlations. This physically motivated design achieves accurate predictions without sacrificing transparency, and the latent coordinates will be analyzed in follow-up work.

This mass catalog enables population analyses that were previously limited by sample size or calibration and will be pursued in forthcoming papers, including tests of continuity equation growth models, duty cycle estimates from the active mass function, consistency checks with the quasar luminosity function via the Eddington ratio distribution, and joint constraints on black hole and galaxy coevolution using host scaling relations.

Due to the limited number of high redshift RM samples and the sparse SDSS spectroscopic coverage at $z>4$, the accuracy of our model is not guaranteed in this regime. As near infrared spectroscopy, such as James Webb telescope, delivers more high $z$ quasars with rest frame optical lines (e.g., H$\beta$, Mg\,\textsc{ii}), we can finetune the network on these new data to improve performance at high redshift, which will enable more reliable mass estimates for $z>4$ quasars.

\bibliographystyle{cas-model2-names}
\bibliography{Quasar-Mass-ML_ads}

\begin{thebibliography}{51}
\expandafter\ifx\csname natexlab\endcsname\relax\def\natexlab#1{#1}\fi
\providecommand{\url}[1]{\texttt{#1}}
\providecommand{\href}[2]{#2}
\providecommand{\path}[1]{#1}
\providecommand{\DOIprefix}{doi:}
\providecommand{\ArXivprefix}{arXiv:}
\providecommand{\URLprefix}{URL: }
\providecommand{\Pubmedprefix}{pmid:}
\providecommand{\doi}[1]{\href{http://dx.doi.org/#1}{\path{#1}}}
\providecommand{\Pubmed}[1]{\href{pmid:#1}{\path{#1}}}
\providecommand{\bibinfo}[2]{#2}
\ifx\xfnm\relax \def\xfnm[#1]{\unskip,\space#1}\fi
\bibitem[{{Alexander} and {Hickox}(2012)}]{2012NewAR..56...93A}
\bibinfo{author}{{Alexander}, D.M.}, \bibinfo{author}{{Hickox}, R.C.}, \bibinfo{year}{2012}.
\newblock \bibinfo{title}{{What drives the growth of black holes?}}
\newblock \bibinfo{journal}{\nar} \bibinfo{volume}{56}, \bibinfo{pages}{93--121}.
\newblock \DOIprefix\doi{10.1016/j.newar.2011.09.003}.
\bibitem[{{Blandford} and {McKee}(1982)}]{1982ApJ...255..419B}
\bibinfo{author}{{Blandford}, R.D.}, \bibinfo{author}{{McKee}, C.F.}, \bibinfo{year}{1982}.
\newblock \bibinfo{title}{{Reverberation mapping of the emission line regions of Seyfert galaxies and quasars.}}
\newblock \bibinfo{journal}{\apj} \bibinfo{volume}{255}, \bibinfo{pages}{419--439}.
\newblock \DOIprefix\doi{10.1086/159843}.
\bibitem[{{Burke} et~al.(2021){Burke}, {Suyu}, {Millon}, {Chen}, {Courbin}, {Refsdal}, {Treu}, {Agnello}, {Mao} and {Meyer}}]{2021Sci...373..789B}
\bibinfo{author}{{Burke}, C.J.}, \bibinfo{author}{{Suyu}, S.H.}, \bibinfo{author}{{Millon}, M.}, \bibinfo{author}{{Chen}, G.C.F.}, \bibinfo{author}{{Courbin}, F.}, \bibinfo{author}{{Refsdal}, S.}, \bibinfo{author}{{Treu}, T.}, \bibinfo{author}{{Agnello}, A.}, \bibinfo{author}{{Mao}, S.}, \bibinfo{author}{{Meyer}, E.T.}, \bibinfo{year}{2021}.
\newblock \bibinfo{title}{{A characteristic optical variability time scale in astrophysical accretion disks}}.
\newblock \bibinfo{journal}{Science} \bibinfo{volume}{373}, \bibinfo{pages}{789--794}.
\newblock \DOIprefix\doi{10.1126/science.aba5344}.
\bibitem[{{Chainakun} et~al.(2022){Chainakun}, {Young}, {Stevens}, {Parker}, {Uttley} and {McHardy}}]{2022MNRAS.513..648C}
\bibinfo{author}{{Chainakun}, P.}, \bibinfo{author}{{Young}, C.}, \bibinfo{author}{{Stevens}, A.L.}, \bibinfo{author}{{Parker}, M.L.}, \bibinfo{author}{{Uttley}, P.}, \bibinfo{author}{{McHardy}, I.M.}, \bibinfo{year}{2022}.
\newblock \bibinfo{title}{{Predicting the black hole mass and accretion rate in X-ray reverberating active galactic nuclei using neural networks}}.
\newblock \bibinfo{journal}{\mnras} \bibinfo{volume}{513}, \bibinfo{pages}{648--661}.
\newblock \DOIprefix\doi{10.1093/mnras/stac1123}.
\bibitem[{{Ciotti} and {Ostriker}(2007)}]{2007ApJ...665.1038C}
\bibinfo{author}{{Ciotti}, L.}, \bibinfo{author}{{Ostriker}, J.P.}, \bibinfo{year}{2007}.
\newblock \bibinfo{title}{{Radiative Feedback from Massive Black Holes in Elliptical Galaxies: AGN Flaring and Central Starburst Fueled by Recycled Gas}}.
\newblock \bibinfo{journal}{\apj} \bibinfo{volume}{665}, \bibinfo{pages}{1038--1056}.
\newblock \DOIprefix\doi{10.1086/519529}.
\bibitem[{{Coatman} et~al.(2017){Coatman}, {Hewett}, {Banerji} and {Richards}}]{2017MNRAS.465.2120C}
\bibinfo{author}{{Coatman}, L.}, \bibinfo{author}{{Hewett}, P.C.}, \bibinfo{author}{{Banerji}, M.}, \bibinfo{author}{{Richards}, G.T.}, \bibinfo{year}{2017}.
\newblock \bibinfo{title}{{Correcting C IV-based virial black hole masses}}.
\newblock \bibinfo{journal}{\mnras} \bibinfo{volume}{465}, \bibinfo{pages}{2120--2142}.
\newblock \DOIprefix\doi{10.1093/mnras/stw2797}.
\bibitem[{{Denney} et~al.(2013){Denney}, {Pogge}, {Assef}, {Barth}, {Bennert}, {Brewer}, {Cid Fernandes}, {Dalla Bont{\'a}}, {De Rosa}, {Dietrich}, {Filippenko}, {Grier}, {Ho}, {Jiang}, {Kaspi}, {Kelly}, {Kim}, {King}, {Kochanek}, {LaMassa}, {Netzer}, {Peterson}, {Rotter}, {Schimoia}, {Shappee}, {Shi}, {Valenti}, {Vestergaard}, {Wang} and {Yoshii}}]{2013ApJ...775...60D}
\bibinfo{author}{{Denney}, K.D.}, \bibinfo{author}{{Pogge}, R.W.}, \bibinfo{author}{{Assef}, R.J.}, \bibinfo{author}{{Barth}, A.J.}, \bibinfo{author}{{Bennert}, V.N.}, \bibinfo{author}{{Brewer}, B.J.}, \bibinfo{author}{{Cid Fernandes}, R.}, \bibinfo{author}{{Dalla Bont{\'a}}, E.}, \bibinfo{author}{{De Rosa}, G.}, \bibinfo{author}{{Dietrich}, M.}, \bibinfo{author}{{Filippenko}, A.V.}, \bibinfo{author}{{Grier}, C.J.}, \bibinfo{author}{{Ho}, L.C.}, \bibinfo{author}{{Jiang}, L.}, \bibinfo{author}{{Kaspi}, S.}, \bibinfo{author}{{Kelly}, B.C.}, \bibinfo{author}{{Kim}, D.C.}, \bibinfo{author}{{King}, A.L.}, \bibinfo{author}{{Kochanek}, C.S.}, \bibinfo{author}{{LaMassa}, S.}, \bibinfo{author}{{Netzer}, H.}, \bibinfo{author}{{Peterson}, B.M.}, \bibinfo{author}{{Rotter}, A.}, \bibinfo{author}{{Schimoia}, J.S.}, \bibinfo{author}{{Shappee}, B.}, \bibinfo{author}{{Shi}, Y.}, \bibinfo{author}{{Valenti}, S.}, \bibinfo{author}{{Vestergaard}, M.}, \bibinfo{author}{{Wang}, J.M.}, \bibinfo{author}{{Yoshii}, Y.},
  \bibinfo{year}{2013}.
\newblock \bibinfo{title}{{C IV line-width anomalies: the perils of low signal-to-noise spectra}}.
\newblock \bibinfo{journal}{\apj} \bibinfo{volume}{775}, \bibinfo{pages}{60}.
\newblock \DOIprefix\doi{10.1088/0004-637X/775/1/60}.
\bibitem[{{Event Horizon Telescope Collaboration} et~al.(2019){Event Horizon Telescope Collaboration}, {Akiyama}, {Alberdi} et~al.}]{2019ApJ...875L...1E}
\bibinfo{author}{{Event Horizon Telescope Collaboration}}, \bibinfo{author}{{Akiyama}, K.}, \bibinfo{author}{{Alberdi}, A.}, et~al., \bibinfo{year}{2019}.
\newblock \bibinfo{title}{{First M87 Event Horizon Telescope Results. I. The Shadow of the Supermassive Black Hole}}.
\newblock \bibinfo{journal}{\apjl} \bibinfo{volume}{875}, \bibinfo{pages}{L1}.
\newblock \DOIprefix\doi{10.3847/2041-8213/ab0ec7}, \href{http://arxiv.org/abs/1906.11238}{\tt arXiv:1906.11238}.
\bibitem[{{Ferrarese} and {Ford}(2005)}]{2005SSRv..116..523F}
\bibinfo{author}{{Ferrarese}, L.}, \bibinfo{author}{{Ford}, H.}, \bibinfo{year}{2005}.
\newblock \bibinfo{title}{{Supermassive Black Holes in Galactic Nuclei: Past, Present and Future Research}}.
\newblock \bibinfo{journal}{\ssr} \bibinfo{volume}{116}, \bibinfo{pages}{523--624}.
\newblock \DOIprefix\doi{10.1007/s11214-005-3947-6}.
\bibitem[{{Ferrarese} and {Merritt}(2000)}]{2000ApJ...539L...9F}
\bibinfo{author}{{Ferrarese}, L.}, \bibinfo{author}{{Merritt}, D.}, \bibinfo{year}{2000}.
\newblock \bibinfo{title}{{A Fundamental Relation Between Supermassive Black Holes and Their Host Galaxies}}.
\newblock \bibinfo{journal}{\apjl} \bibinfo{volume}{539}, \bibinfo{pages}{L9--L12}.
\newblock \DOIprefix\doi{10.1086/312838}.
\bibitem[{{Graham} and {Scott}(2013)}]{2013ApJ...764..151G}
\bibinfo{author}{{Graham}, A.W.}, \bibinfo{author}{{Scott}, N.}, \bibinfo{year}{2013}.
\newblock \bibinfo{title}{{The M $_{BH}$-L $_{spheroid}$ Relation at High and Low Masses, the Quadratic Growth of Black Holes, and Intermediate-mass Black Hole Candidates}}.
\newblock \bibinfo{journal}{\apj} \bibinfo{volume}{764}, \bibinfo{pages}{151}.
\newblock \DOIprefix\doi{10.1088/0004-637X/764/2/151}.
\bibitem[{{Greene} and {Ho}(2007)}]{2007ApJ...667..131G}
\bibinfo{author}{{Greene}, J.E.}, \bibinfo{author}{{Ho}, L.C.}, \bibinfo{year}{2007}.
\newblock \bibinfo{title}{{The Mass Function of Active Black Holes in the Local Universe}}.
\newblock \bibinfo{journal}{\apj} \bibinfo{volume}{667}, \bibinfo{pages}{131--148}.
\newblock \DOIprefix\doi{10.1086/520497}, \href{http://arxiv.org/abs/0705.0020}{\tt arXiv:0705.0020}.
\bibitem[{{H{\"a}ring} and {Rix}(2004)}]{2004ApJ...604L..89H}
\bibinfo{author}{{H{\"a}ring}, N.}, \bibinfo{author}{{Rix}, H.W.}, \bibinfo{year}{2004}.
\newblock \bibinfo{title}{{On the Black Hole Mass-Bulge Mass Relation}}.
\newblock \bibinfo{journal}{\apjl} \bibinfo{volume}{604}, \bibinfo{pages}{L89--L92}.
\newblock \DOIprefix\doi{10.1086/383567}.
\bibitem[{{He} et~al.(2022){He}, {Qin}, {Wang} and {Wu}}]{2022RAA....22h5014H}
\bibinfo{author}{{He}, D.}, \bibinfo{author}{{Qin}, Y.}, \bibinfo{author}{{Wang}, F.Y.}, \bibinfo{author}{{Wu}, X.F.}, \bibinfo{year}{2022}.
\newblock \bibinfo{title}{{Predicting Supermassive Black Hole Masses Using Machine Learning}}.
\newblock \bibinfo{journal}{Research in Astronomy and Astrophysics} \bibinfo{volume}{22}, \bibinfo{pages}{085014}.
\newblock \DOIprefix\doi{10.1088/1674-4527/ac8190}.
\bibitem[{{Hopkins} et~al.(2008){Hopkins}, {Hernquist}, {Cox}, {Kere{\v{s}}}, {Aravena}, {Bernardi}, {Blanton}, {Brinchmann}, {Bundy}, {Busha}, {Caldwell}, {Dalcanton}, {Davis}, {de Bruyn}, {Dekel}, {Egami}, {Finkbeiner}, {Giovanelli}, {Gnedin}, {Gottl{"o}ber}, {Greve}, {Grogin}, {Haiman}, {Iono}, {Jarvis}, {Kravtsov}, {Lahav}, {Loeb}, {Lutz}, {McDonald}, {Murray}, {Norberg}, {Ostriker}, {Padmanabhan}, {Papovich}, {Pierce}, {Salvaterra}, {Scannapieco}, {Schmidt}, {Shapley}, {Sheth}, {Stewart}, {Strauss}, {Toft}, {Wetzel}, {Wyithe} and {Zabludoff}}]{2008ApJS..175..356H}
\bibinfo{author}{{Hopkins}, P.F.}, \bibinfo{author}{{Hernquist}, L.}, \bibinfo{author}{{Cox}, T.J.}, \bibinfo{author}{{Kere{\v{s}}}, D.}, \bibinfo{author}{{Aravena}, M.}, \bibinfo{author}{{Bernardi}, M.}, \bibinfo{author}{{Blanton}, M.}, \bibinfo{author}{{Brinchmann}, J.}, \bibinfo{author}{{Bundy}, K.}, \bibinfo{author}{{Busha}, M.T.}, \bibinfo{author}{{Caldwell}, R.}, \bibinfo{author}{{Dalcanton}, J.}, \bibinfo{author}{{Davis}, M.}, \bibinfo{author}{{de Bruyn}, A.G.}, \bibinfo{author}{{Dekel}, A.}, \bibinfo{author}{{Egami}, E.}, \bibinfo{author}{{Finkbeiner}, D.}, \bibinfo{author}{{Giovanelli}, R.}, \bibinfo{author}{{Gnedin}, N.Y.}, \bibinfo{author}{{Gottl{"o}ber}, S.}, \bibinfo{author}{{Greve}, T.R.}, \bibinfo{author}{{Grogin}, N.A.}, \bibinfo{author}{{Haiman}, Z.}, \bibinfo{author}{{Iono}, D.}, \bibinfo{author}{{Jarvis}, M.}, \bibinfo{author}{{Kravtsov}, A.V.}, \bibinfo{author}{{Lahav}, O.}, \bibinfo{author}{{Loeb}, A.}, \bibinfo{author}{{Lutz}, D.}, \bibinfo{author}{{McDonald}, P.},
  \bibinfo{author}{{Murray}, N.}, \bibinfo{author}{{Norberg}, P.}, \bibinfo{author}{{Ostriker}, J.P.}, \bibinfo{author}{{Padmanabhan}, T.}, \bibinfo{author}{{Papovich}, C.}, \bibinfo{author}{{Pierce}, C.M.}, \bibinfo{author}{{Salvaterra}, R.}, \bibinfo{author}{{Scannapieco}, E.}, \bibinfo{author}{{Schmidt}, B.P.}, \bibinfo{author}{{Shapley}, A.}, \bibinfo{author}{{Sheth}, R.K.}, \bibinfo{author}{{Stewart}, K.R.}, \bibinfo{author}{{Strauss}, M.A.}, \bibinfo{author}{{Toft}, S.}, \bibinfo{author}{{Wetzel}, A.}, \bibinfo{author}{{Wyithe}, J.S.B.}, \bibinfo{author}{{Zabludoff}, A.I.}, \bibinfo{year}{2008}.
\newblock \bibinfo{title}{{A Cosmological Framework for the Co-evolution of Quasars, Supermassive Black Holes, and Elliptical Galaxies. I. Galaxy Mergers and Quasar Activity}}.
\newblock \bibinfo{journal}{\apjs} \bibinfo{volume}{175}, \bibinfo{pages}{356--389}.
\newblock \DOIprefix\doi{10.1086/524362}.
\bibitem[{{Inayoshi} et~al.(2020){Inayoshi}, {Visbal} and {Haiman}}]{2020ARA&A..58...27I}
\bibinfo{author}{{Inayoshi}, K.}, \bibinfo{author}{{Visbal}, E.}, \bibinfo{author}{{Haiman}, Z.}, \bibinfo{year}{2020}.
\newblock \bibinfo{title}{{The Assembly of the First Massive Black Holes}}.
\newblock \bibinfo{journal}{\araa} \bibinfo{volume}{58}, \bibinfo{pages}{27--97}.
\newblock \DOIprefix\doi{10.1146/annurev-astro-120419-014455}.
\bibitem[{{Ivezi{\'c}} et~al.(2019){Ivezi{\'c}}, {Kahn}, {Tyson}, {Abel}, {Acosta}, {Allsman}, {AlSayyad}, {Anderson}, {Andrew} and {Angeli}}]{2019ApJ...873..111I}
\bibinfo{author}{{Ivezi{\'c}}, {\v{Z}}.}, \bibinfo{author}{{Kahn}, S.M.}, \bibinfo{author}{{Tyson}, J.A.}, \bibinfo{author}{{Abel}, B.}, \bibinfo{author}{{Acosta}, E.}, \bibinfo{author}{{Allsman}, R.}, \bibinfo{author}{{AlSayyad}, Y.}, \bibinfo{author}{{Anderson}, S.F.}, \bibinfo{author}{{Andrew}, J.}, \bibinfo{author}{{Angeli}, G.Z.e.a.}, \bibinfo{year}{2019}.
\newblock \bibinfo{title}{{LSST: From Science Drivers to Reference Design and Anticipated Data Products}}.
\newblock \bibinfo{journal}{\apj} \bibinfo{volume}{873}, \bibinfo{pages}{111}.
\newblock \DOIprefix\doi{10.3847/1538-4357/ab042c}.
\bibitem[{{Izumi} et~al.(2019){Izumi}, {Onoue}, {Fujimoto}, {Ueda}, {Matsuoka}, {Nagao}, {Kohno}, {Toba}, {Shirakata}, {Nakanishi}, {Yamaguchi}, {Akiyama}, {Alexander}, {Asami}, {Bruno}, {Furusawa}, {Goto}, {Ishikawa}, {Kubo}, {Lee}, {Lu}, {Mawatari}, {Meyer}, {Ni}, {Ono}, {Shibuya}, {Taniguchi}, {Tanaka} and {Wang}}]{2019PASJ...71..111I}
\bibinfo{author}{{Izumi}, T.}, \bibinfo{author}{{Onoue}, M.}, \bibinfo{author}{{Fujimoto}, S.}, \bibinfo{author}{{Ueda}, Y.}, \bibinfo{author}{{Matsuoka}, Y.}, \bibinfo{author}{{Nagao}, T.}, \bibinfo{author}{{Kohno}, K.}, \bibinfo{author}{{Toba}, Y.}, \bibinfo{author}{{Shirakata}, H.}, \bibinfo{author}{{Nakanishi}, K.}, \bibinfo{author}{{Yamaguchi}, Y.}, \bibinfo{author}{{Akiyama}, M.}, \bibinfo{author}{{Alexander}, D.M.}, \bibinfo{author}{{Asami}, N.}, \bibinfo{author}{{Bruno}, G.}, \bibinfo{author}{{Furusawa}, H.}, \bibinfo{author}{{Goto}, T.}, \bibinfo{author}{{Ishikawa}, S.}, \bibinfo{author}{{Kubo}, M.}, \bibinfo{author}{{Lee}, K.S.}, \bibinfo{author}{{Lu}, T.M.}, \bibinfo{author}{{Mawatari}, K.}, \bibinfo{author}{{Meyer}, R.A.}, \bibinfo{author}{{Ni}, Y.}, \bibinfo{author}{{Ono}, Y.}, \bibinfo{author}{{Shibuya}, T.}, \bibinfo{author}{{Taniguchi}, Y.}, \bibinfo{author}{{Tanaka}, M.}, \bibinfo{author}{{Wang}, F.}, \bibinfo{year}{2019}.
\newblock \bibinfo{title}{{Subaru High-z Exploration of Low-Luminosity Quasars (SHELLQs) Project. VIII. Black Hole Masses, Accretion Rates, and Narrow Emission-Line Diagnostics for Quasars at z \textasciitilde 6}}.
\newblock \bibinfo{journal}{\pasj} \bibinfo{volume}{71}, \bibinfo{pages}{111}.
\newblock \DOIprefix\doi{10.1093/pasj/psz109}.
\bibitem[{{Jin} and {Davis}(2023)}]{2023arXiv231019406J}
\bibinfo{author}{{Jin}, Z.}, \bibinfo{author}{{Davis}, B.L.}, \bibinfo{year}{2023}.
\newblock \bibinfo{title}{{Discovering black hole mass scaling relations from multi-wavelength AGN observations with symbolic regression}}.
\newblock \bibinfo{journal}{arXiv e-prints} \href{http://arxiv.org/abs/2310.19406}{\tt arXiv:2310.19406}.
\bibitem[{{Kelly} and {Shen}(2013)}]{2013ApJ...764...45K}
\bibinfo{author}{{Kelly}, B.C.}, \bibinfo{author}{{Shen}, Y.}, \bibinfo{year}{2013}.
\newblock \bibinfo{title}{{The Demographics of Broad-line Quasars in the Mass-Luminosity Plane. II. Black Hole Mass Function and Eddington Ratio Distribution}}.
\newblock \bibinfo{journal}{\apj} \bibinfo{volume}{764}, \bibinfo{pages}{45}.
\newblock \DOIprefix\doi{10.1088/0004-637X/764/1/45}.
\bibitem[{{Kollmeier} et~al.(2006){Kollmeier}, {Onken}, {Kochanek}, {Gould}, {Weinberg}, {Dietrich}, {Cool}, {Dey}, {Eisenstein}, {Jannuzi}, {Le Floc'h} and {Stern}}]{2006ApJ...648..128K}
\bibinfo{author}{{Kollmeier}, J.A.}, \bibinfo{author}{{Onken}, C.A.}, \bibinfo{author}{{Kochanek}, C.S.}, \bibinfo{author}{{Gould}, A.}, \bibinfo{author}{{Weinberg}, D.H.}, \bibinfo{author}{{Dietrich}, M.}, \bibinfo{author}{{Cool}, R.}, \bibinfo{author}{{Dey}, A.}, \bibinfo{author}{{Eisenstein}, D.J.}, \bibinfo{author}{{Jannuzi}, B.T.}, \bibinfo{author}{{Le Floc'h}, E.}, \bibinfo{author}{{Stern}, D.}, \bibinfo{year}{2006}.
\newblock \bibinfo{title}{{Black Hole Masses and Eddington Ratios at 0.3 < z < 4}}.
\newblock \bibinfo{journal}{\apj} \bibinfo{volume}{648}, \bibinfo{pages}{128--139}.
\newblock \DOIprefix\doi{10.1086/505646}, \href{http://arxiv.org/abs/astro-ph/0508657}{\tt arXiv:astro-ph/0508657}.
\bibitem[{{Kormendy} and {Ho}(2013)}]{2013ARA&A..51..511K}
\bibinfo{author}{{Kormendy}, J.}, \bibinfo{author}{{Ho}, L.C.}, \bibinfo{year}{2013}.
\newblock \bibinfo{title}{{Coevolution (or Not) of Supermassive Black Holes and Host Galaxies}}.
\newblock \bibinfo{journal}{\araa} \bibinfo{volume}{51}, \bibinfo{pages}{511--653}.
\newblock \DOIprefix\doi{10.1146/annurev-astro-082708-101811}.
\bibitem[{{Kormendy} and {Richstone}(1995)}]{1995ARA&A..33..581K}
\bibinfo{author}{{Kormendy}, J.}, \bibinfo{author}{{Richstone}, D.}, \bibinfo{year}{1995}.
\newblock \bibinfo{title}{{Inward Bound---The Search For Supermassive Black Holes In Galactic Nuclei}}.
\newblock \bibinfo{journal}{\araa} \bibinfo{volume}{33}, \bibinfo{pages}{581--624}.
\newblock \DOIprefix\doi{10.1146/annurev.aa.33.090195.003053}.
\bibitem[{{Kuo} et~al.(2011){Kuo}, {Braatz}, {Condon}, {Impellizzeri}, {Lo}, {Zaw}, {Schenker}, {Henkel}, {Reid} and {Greene}}]{2011ApJ...727...20K}
\bibinfo{author}{{Kuo}, C.Y.}, \bibinfo{author}{{Braatz}, J.A.}, \bibinfo{author}{{Condon}, J.J.}, \bibinfo{author}{{Impellizzeri}, C.M.V.}, \bibinfo{author}{{Lo}, K.Y.}, \bibinfo{author}{{Zaw}, I.}, \bibinfo{author}{{Schenker}, M.}, \bibinfo{author}{{Henkel}, C.}, \bibinfo{author}{{Reid}, M.J.}, \bibinfo{author}{{Greene}, J.E.}, \bibinfo{year}{2011}.
\newblock \bibinfo{title}{{The Megamaser Cosmology Project. III. Accurate Masses of Seven Supermassive Black Holes in Active Galaxies with Circumnuclear Megamaser Disks}}.
\newblock \bibinfo{journal}{\apj} \bibinfo{volume}{727}, \bibinfo{pages}{20}.
\newblock \DOIprefix\doi{10.1088/0004-637X/727/1/20}, \href{http://arxiv.org/abs/1008.2146}{\tt arXiv:1008.2146}.
\bibitem[{{Lin} et~al.(2023){Lin}, {Leung}, {Shu}, {Song}, {Pei}, {Fu}, {Peng}, {Ye} and {Liu}}]{2023MNRAS.518.4921L}
\bibinfo{author}{{Lin}, H.S.}, \bibinfo{author}{{Leung}, T.C.}, \bibinfo{author}{{Shu}, F.}, \bibinfo{author}{{Song}, H.}, \bibinfo{author}{{Pei}, L.}, \bibinfo{author}{{Fu}, H.Q.}, \bibinfo{author}{{Peng}, C.}, \bibinfo{author}{{Ye}, J.L.}, \bibinfo{author}{{Liu}, X.}, \bibinfo{year}{2023}.
\newblock \bibinfo{title}{{AGNet: An Attention-Based Neural Network to Predict Supermassive Black Hole Masses from Astronomical Light Curves}}.
\newblock \bibinfo{journal}{\mnras} \bibinfo{volume}{518}, \bibinfo{pages}{4921--4933}.
\newblock \DOIprefix\doi{10.1093/mnras/stac3394}.
\bibitem[{{Lyke} et~al.(2020){Lyke}, {Higley}, {McLane}, {Schurhammer}, {Myers}, {Ross}, {Dawson}, {Chabanier}, {Martini}, {Busca}, {du Mas des Bourboux}, {Salvato}, {Streblyanska}, {Zarrouk}, {Burtin}, {Anderson}, {Bautista}, {Bizyaev}, {Brandt}, {Brinkmann}, {Brownstein}, {Comparat}, {Green}, {de la Macorra}, {Mu{\~n}oz Guti{\'e}rrez}, {Hou}, {Newman}, {Palanque-Delabrouille}, {P{\^a}ris}, {Percival}, {Petitjean}, {Rich}, {Rossi}, {Schneider}, {Smith}, {Vivek} and {Weaver}}]{2020ApJS..250....8L}
\bibinfo{author}{{Lyke}, B.W.}, \bibinfo{author}{{Higley}, A.N.}, \bibinfo{author}{{McLane}, J.N.}, \bibinfo{author}{{Schurhammer}, D.P.}, \bibinfo{author}{{Myers}, A.D.}, \bibinfo{author}{{Ross}, A.J.}, \bibinfo{author}{{Dawson}, K.}, \bibinfo{author}{{Chabanier}, S.}, \bibinfo{author}{{Martini}, P.}, \bibinfo{author}{{Busca}, N.G.}, \bibinfo{author}{{du Mas des Bourboux}, H.}, \bibinfo{author}{{Salvato}, M.}, \bibinfo{author}{{Streblyanska}, A.}, \bibinfo{author}{{Zarrouk}, P.}, \bibinfo{author}{{Burtin}, E.}, \bibinfo{author}{{Anderson}, S.F.}, \bibinfo{author}{{Bautista}, J.}, \bibinfo{author}{{Bizyaev}, D.}, \bibinfo{author}{{Brandt}, W.N.}, \bibinfo{author}{{Brinkmann}, J.}, \bibinfo{author}{{Brownstein}, J.R.}, \bibinfo{author}{{Comparat}, J.}, \bibinfo{author}{{Green}, P.}, \bibinfo{author}{{de la Macorra}, A.}, \bibinfo{author}{{Mu{\~n}oz Guti{\'e}rrez}, A.}, \bibinfo{author}{{Hou}, J.}, \bibinfo{author}{{Newman}, J.A.}, \bibinfo{author}{{Palanque-Delabrouille}, N.}, \bibinfo{author}{{P{\^a}ris},
  I.}, \bibinfo{author}{{Percival}, W.J.}, \bibinfo{author}{{Petitjean}, P.}, \bibinfo{author}{{Rich}, J.}, \bibinfo{author}{{Rossi}, G.}, \bibinfo{author}{{Schneider}, D.P.}, \bibinfo{author}{{Smith}, A.}, \bibinfo{author}{{Vivek}, M.}, \bibinfo{author}{{Weaver}, B.A.}, \bibinfo{year}{2020}.
\newblock \bibinfo{title}{{The Sloan Digital Sky Survey Quasar Catalog: Sixteenth Data Release}}.
\newblock \bibinfo{journal}{\apjs} \bibinfo{volume}{250}, \bibinfo{pages}{8}.
\newblock \DOIprefix\doi{10.3847/1538-4365/aba623}, \href{http://arxiv.org/abs/2007.09001}{\tt arXiv:2007.09001}.
\bibitem[{{Matthews} et~al.(2023){Matthews}, {Strong-Wright}, {Knigge}, {Hewett}, {Temple}, {Long}, {Rankine}, {Stepney}, {Banerji} and {Richards}}]{2023MNRAS.526.3967M}
\bibinfo{author}{{Matthews}, J.H.}, \bibinfo{author}{{Strong-Wright}, J.}, \bibinfo{author}{{Knigge}, C.}, \bibinfo{author}{{Hewett}, P.}, \bibinfo{author}{{Temple}, M.J.}, \bibinfo{author}{{Long}, K.S.}, \bibinfo{author}{{Rankine}, A.L.}, \bibinfo{author}{{Stepney}, M.}, \bibinfo{author}{{Banerji}, M.}, \bibinfo{author}{{Richards}, G.T.}, \bibinfo{year}{2023}.
\newblock \bibinfo{title}{{A disc wind model for blueshifts in quasar broad emission lines}}.
\newblock \bibinfo{journal}{\mnras} \bibinfo{volume}{526}, \bibinfo{pages}{3967--3986}.
\newblock \DOIprefix\doi{10.1093/mnras/stad2895}.
\bibitem[{{McConnell} and {Ma}(2013)}]{2013ApJ...764..184M}
\bibinfo{author}{{McConnell}, N.J.}, \bibinfo{author}{{Ma}, C.P.}, \bibinfo{year}{2013}.
\newblock \bibinfo{title}{{Revisiting the Scaling Relations of Black Hole Masses and Host Galaxy Properties}}.
\newblock \bibinfo{journal}{\apj} \bibinfo{volume}{764}, \bibinfo{pages}{184}.
\newblock \DOIprefix\doi{10.1088/0004-637X/764/2/184}.
\bibitem[{{McHardy} et~al.(2006){McHardy}, {Koerding}, {Knigge}, {Uttley} and {Fender}}]{2006Natur.444..730M}
\bibinfo{author}{{McHardy}, I.M.}, \bibinfo{author}{{Koerding}, E.}, \bibinfo{author}{{Knigge}, C.}, \bibinfo{author}{{Uttley}, P.}, \bibinfo{author}{{Fender}, R.P.}, \bibinfo{year}{2006}.
\newblock \bibinfo{title}{{Active galactic nuclei as scaled-up Galactic black holes}}.
\newblock \bibinfo{journal}{\nat} \bibinfo{volume}{444}, \bibinfo{pages}{730--732}.
\newblock \DOIprefix\doi{10.1038/nature05389}.
\bibitem[{{McInnes} et~al.(2018){McInnes}, {Healy} and {Melville}}]{2018arXiv180203426M}
\bibinfo{author}{{McInnes}, L.}, \bibinfo{author}{{Healy}, J.}, \bibinfo{author}{{Melville}, J.}, \bibinfo{year}{2018}.
\newblock \bibinfo{title}{{UMAP: Uniform Manifold Approximation and Projection for Dimension Reduction}}.
\newblock \bibinfo{journal}{arXiv e-prints} \href{http://arxiv.org/abs/1802.03426}{\tt arXiv:1802.03426}.
\bibitem[{{Mej{\'\i}a-Restrepo} et~al.(2017){Mej{\'\i}a-Restrepo}, {Lira}, {Netzer}, {Trakhtenbrot} and {Capellupo}}]{2017FrASS...4...70M}
\bibinfo{author}{{Mej{\'\i}a-Restrepo}, J.}, \bibinfo{author}{{Lira}, P.}, \bibinfo{author}{{Netzer}, H.}, \bibinfo{author}{{Trakhtenbrot}, B.}, \bibinfo{author}{{Capellupo}, D.}, \bibinfo{year}{2017}.
\newblock \bibinfo{title}{{The virial factor and biases in single epoch black hole mass determinations}}.
\newblock \bibinfo{journal}{Frontiers in Astronomy and Space Sciences} \bibinfo{volume}{4}, \bibinfo{pages}{70}.
\newblock \DOIprefix\doi{10.3389/fspas.2017.00070}.
\bibitem[{Mej{\'\i}a-Restrepo et~al.(2018)Mej{\'\i}a-Restrepo, Trakhtenbrot, Lira, Netzer and Capellupo}]{2018MNRAS.478.1929M}
\bibinfo{author}{Mej{\'\i}a-Restrepo, J.E.}, \bibinfo{author}{Trakhtenbrot, B.}, \bibinfo{author}{Lira, P.}, \bibinfo{author}{Netzer, H.}, \bibinfo{author}{Capellupo, D.M.}, \bibinfo{year}{2018}.
\newblock \bibinfo{title}{{The reliability of C IV $\lambda$ 1549 broad-line as a virial black hole mass estimator for active galactic nuclei}}.
\newblock \bibinfo{journal}{Monthly Notices of the Royal Astronomical Society} \bibinfo{volume}{478}, \bibinfo{pages}{1929--1952}.
\newblock \DOIprefix\doi{10.1093/mnras/sty1153}.
\bibitem[{{Merritt} and {Ferrarese}(2001)}]{2001ApJ...547..140M}
\bibinfo{author}{{Merritt}, D.}, \bibinfo{author}{{Ferrarese}, L.}, \bibinfo{year}{2001}.
\newblock \bibinfo{title}{{The M\_{bh}-sigma Relation for Supermassive Black Holes}}.
\newblock \bibinfo{journal}{\apj} \bibinfo{volume}{547}, \bibinfo{pages}{140--145}.
\newblock \DOIprefix\doi{10.1086/318372}.
\bibitem[{{Peterson}(1993)}]{1993PASP..105..247P}
\bibinfo{author}{{Peterson}, B.M.}, \bibinfo{year}{1993}.
\newblock \bibinfo{title}{{Reverberation Mapping of Active Galactic Nuclei}}.
\newblock \bibinfo{journal}{\pasp} \bibinfo{volume}{105}, \bibinfo{pages}{247}.
\newblock \DOIprefix\doi{10.1086/133140}.
\bibitem[{{Richards} et~al.(2002){Richards}, {Vanden Berk}, {Reichard}, {Hall}, {Schneider}, {SubbaRao}, {Thakar}, {York} and {Strateva}}]{2002AJ....124....1R}
\bibinfo{author}{{Richards}, G.T.}, \bibinfo{author}{{Vanden Berk}, D.E.}, \bibinfo{author}{{Reichard}, T.A.}, \bibinfo{author}{{Hall}, P.B.}, \bibinfo{author}{{Schneider}, D.P.}, \bibinfo{author}{{SubbaRao}, M.}, \bibinfo{author}{{Thakar}, A.R.}, \bibinfo{author}{{York}, D.G.}, \bibinfo{author}{{Strateva}, I.V.}, \bibinfo{year}{2002}.
\newblock \bibinfo{title}{{Broad Emission-Line Shifts in Quasars: An Orientation Measure for Quasars?}}
\newblock \bibinfo{journal}{\aj} \bibinfo{volume}{124}, \bibinfo{pages}{1--17}.
\newblock \DOIprefix\doi{10.1086/339056}.
\bibitem[{{Saglia} et~al.(2016){Saglia}, {Opitsch}, {Erwin}, {Thomas}, {Beifiori}, {Fabricius}, {F{"o}rster Schreiber}, {Mendel} and {Bender}}]{2016ApJ...818...47S}
\bibinfo{author}{{Saglia}, R.P.}, \bibinfo{author}{{Opitsch}, M.}, \bibinfo{author}{{Erwin}, P.}, \bibinfo{author}{{Thomas}, J.}, \bibinfo{author}{{Beifiori}, A.}, \bibinfo{author}{{Fabricius}, M.}, \bibinfo{author}{{F{"o}rster Schreiber}, N.M.}, \bibinfo{author}{{Mendel}, J.T.}, \bibinfo{author}{{Bender}, R.}, \bibinfo{year}{2016}.
\newblock \bibinfo{title}{{The SINFONI Black Hole Survey: The Black Hole Fundamental Plane Revisited and the Paths of (Co)evolution of Supermassive Black Holes and Bulges}}.
\newblock \bibinfo{journal}{\apj} \bibinfo{volume}{818}, \bibinfo{pages}{47}.
\newblock \DOIprefix\doi{10.3847/0004-637X/818/1/47}.
\bibitem[{{Schawinski} et~al.(2010){Schawinski}, {Dowlin}, {Thomas}, {Urry} and {Edmondson}}]{2010ApJ...711..284S}
\bibinfo{author}{{Schawinski}, K.}, \bibinfo{author}{{Dowlin}, N.}, \bibinfo{author}{{Thomas}, D.}, \bibinfo{author}{{Urry}, C.M.}, \bibinfo{author}{{Edmondson}, E.I.}, \bibinfo{year}{2010}.
\newblock \bibinfo{title}{{The Fundamentally Different Co-Evolution of Supermassive Black Holes and Their Host Galaxies}}.
\newblock \bibinfo{journal}{\apj} \bibinfo{volume}{711}, \bibinfo{pages}{284--302}.
\newblock \DOIprefix\doi{10.1088/0004-637X/711/1/284}.
\bibitem[{{Schechter}(1976)}]{1976ApJ...203..297S}
\bibinfo{author}{{Schechter}, P.}, \bibinfo{year}{1976}.
\newblock \bibinfo{title}{{An analytic expression for the luminosity function for galaxies.}}
\newblock \bibinfo{journal}{\apj} \bibinfo{volume}{203}, \bibinfo{pages}{297--306}.
\newblock \DOIprefix\doi{10.1086/154079}.
\bibitem[{{Schramm} and {Silverman}(2013)}]{2013ApJ...767...13S}
\bibinfo{author}{{Schramm}, M.}, \bibinfo{author}{{Silverman}, J.D.}, \bibinfo{year}{2013}.
\newblock \bibinfo{title}{{The Black Hole-Bulge Mass Relation of Active Galactic Nuclei in the Extended Chandra Deep Field-South Survey}}.
\newblock \bibinfo{journal}{\apj} \bibinfo{volume}{767}, \bibinfo{pages}{13}.
\newblock \DOIprefix\doi{10.1088/0004-637X/767/1/13}.
\bibitem[{{Schulze} and {Wisotzki}(2010)}]{2010A&A...516A..87S}
\bibinfo{author}{{Schulze}, A.}, \bibinfo{author}{{Wisotzki}, L.}, \bibinfo{year}{2010}.
\newblock \bibinfo{title}{{Low redshift AGN in the Hamburg/ESO Survey . II. The active black hole mass function and the distribution function of Eddington ratios}}.
\newblock \bibinfo{journal}{\aap} \bibinfo{volume}{516}, \bibinfo{pages}{A87}.
\newblock \DOIprefix\doi{10.1051/0004-6361/201014193}, \href{http://arxiv.org/abs/1004.2671}{\tt arXiv:1004.2671}.
\bibitem[{{Shankar} et~al.(2019){Shankar}, {Bernardi}, {Richardson}, {Lang}, {Xie}, {Koss}, {Krishnan}, {Eracleous} and {Colpi}}]{2019MNRAS.485.1278S}
\bibinfo{author}{{Shankar}, F.}, \bibinfo{author}{{Bernardi}, M.}, \bibinfo{author}{{Richardson}, S.N.}, \bibinfo{author}{{Lang}, P.}, \bibinfo{author}{{Xie}, M.}, \bibinfo{author}{{Koss}, M.J.}, \bibinfo{author}{{Krishnan}, C.}, \bibinfo{author}{{Eracleous}, M.}, \bibinfo{author}{{Colpi}, M.}, \bibinfo{year}{2019}.
\newblock \bibinfo{title}{{Black hole scaling relations of active and quiescent galaxies: addressing selection effects and constraining virial factors}}.
\newblock \bibinfo{journal}{\mnras} \bibinfo{volume}{485}, \bibinfo{pages}{1278--1300}.
\newblock \DOIprefix\doi{10.1093/mnras/stz509}.
\bibitem[{{Shankar} et~al.(2016){Shankar}, {Bernardi}, {Sheth}, {Ferrarese}, {Graham}, {Marconi}, {Savorgnan}, {La Barbera}, {Lasky} and {Haigh}}]{2016MNRAS.460.3119S}
\bibinfo{author}{{Shankar}, F.}, \bibinfo{author}{{Bernardi}, M.}, \bibinfo{author}{{Sheth}, R.K.}, \bibinfo{author}{{Ferrarese}, L.}, \bibinfo{author}{{Graham}, A.W.}, \bibinfo{author}{{Marconi}, A.}, \bibinfo{author}{{Savorgnan}, G.}, \bibinfo{author}{{La Barbera}, F.}, \bibinfo{author}{{Lasky}, P.D.}, \bibinfo{author}{{Haigh}, I.P.}, \bibinfo{year}{2016}.
\newblock \bibinfo{title}{{Selection bias in dynamically measured supermassive black hole samples: its consequences and the quest for the most fundamental relation}}.
\newblock \bibinfo{journal}{\mnras} \bibinfo{volume}{460}, \bibinfo{pages}{3119--3144}.
\newblock \DOIprefix\doi{10.1093/mnras/stw803}.
\bibitem[{{Shen} et~al.(2015){Shen}, {Brandt}, {Dawson}, {Hall}, {MacLeod}, {Richards}, {Schneider}, {Trump}, {Byun}, {Chen} et~al.}]{2015ApJS..216....4S}
\bibinfo{author}{{Shen}, Y.}, \bibinfo{author}{{Brandt}, W.N.}, \bibinfo{author}{{Dawson}, K.S.}, \bibinfo{author}{{Hall}, P.B.}, \bibinfo{author}{{MacLeod}, C.L.}, \bibinfo{author}{{Richards}, G.T.}, \bibinfo{author}{{Schneider}, D.P.}, \bibinfo{author}{{Trump}, J.R.}, \bibinfo{author}{{Byun}, D.Y.}, \bibinfo{author}{{Chen}, Y.}, et~al., \bibinfo{year}{2015}.
\newblock \bibinfo{title}{{The Sloan Digital Sky Survey Reverberation Mapping Project: Technical Implementation and Sample Characterization}}.
\newblock \bibinfo{journal}{\apjs} \bibinfo{volume}{216}, \bibinfo{pages}{4}.
\newblock \DOIprefix\doi{10.1088/0067-0049/216/1/4}.
\bibitem[{{Shen} et~al.(2008){Shen}, {Greene}, {Strauss}, {Richards}, {Schneider}, {McDonald}, {Bailey}, {Fan}, {Hall}, {Jiang}, {Ross}, {Brandt}, {Kurtz} and {Vanden Berk}}]{2008ApJ...680..169S}
\bibinfo{author}{{Shen}, Y.}, \bibinfo{author}{{Greene}, J.E.}, \bibinfo{author}{{Strauss}, M.A.}, \bibinfo{author}{{Richards}, G.T.}, \bibinfo{author}{{Schneider}, D.P.}, \bibinfo{author}{{McDonald}, P.}, \bibinfo{author}{{Bailey}, S.}, \bibinfo{author}{{Fan}, X.}, \bibinfo{author}{{Hall}, P.B.}, \bibinfo{author}{{Jiang}, L.}, \bibinfo{author}{{Ross}, N.P.}, \bibinfo{author}{{Brandt}, W.N.}, \bibinfo{author}{{Kurtz}, M.J.}, \bibinfo{author}{{Vanden Berk}, D.E.}, \bibinfo{year}{2008}.
\newblock \bibinfo{title}{{Biases in virial black hole masses: an SDSS perspective}}.
\newblock \bibinfo{journal}{\apj} \bibinfo{volume}{680}, \bibinfo{pages}{169--190}.
\newblock \DOIprefix\doi{10.1086/587475}.
\bibitem[{Shen et~al.(2024)Shen, Grier, Horne, Stone, Li, Yang, Homayouni, Trump, Anderson, Brandt, Hall, Ho, Jiang, Petitjean, Schneider, Tao, Donnan, AlSayyad, Bershady, ... and Zou}]{shen2024sloan}
\bibinfo{author}{Shen, Y.}, \bibinfo{author}{Grier, C.J.}, \bibinfo{author}{Horne, K.}, \bibinfo{author}{Stone, Z.}, \bibinfo{author}{Li, J.I.}, \bibinfo{author}{Yang, Q.}, \bibinfo{author}{Homayouni, Y.}, \bibinfo{author}{Trump, J.R.}, \bibinfo{author}{Anderson, S.F.}, \bibinfo{author}{Brandt, W.N.}, \bibinfo{author}{Hall, P.B.}, \bibinfo{author}{Ho, L.C.}, \bibinfo{author}{Jiang, L.}, \bibinfo{author}{Petitjean, P.}, \bibinfo{author}{Schneider, D.P.}, \bibinfo{author}{Tao, C.}, \bibinfo{author}{Donnan, F.R.}, \bibinfo{author}{AlSayyad, Y.}, \bibinfo{author}{Bershady, M.A.}, \bibinfo{author}{...}, \bibinfo{author}{Zou, H.}, \bibinfo{year}{2024}.
\newblock \bibinfo{title}{The sloan digital sky survey reverberation mapping project: Key results}.
\newblock \bibinfo{journal}{The Astrophysical Journal Supplement Series} \bibinfo{volume}{272}, \bibinfo{pages}{26}.
\newblock \DOIprefix\doi{10.3847/1538-4365/ad3936}.
\bibitem[{{Shen} et~al.(2011){Shen}, {Richards}, {Strauss}, {Hall}, {Schneider}, {Snedden}, {Bizyaev}, {Brewington}, {Brinkmann}, {Eisenstein}, {Frieman}, {Gray}, {Gunn}, {Hao}, {Harrington}, {Harvanek}, {Kent}, {Kleinman}, {Kniazev}, {Krzesinski}, {Long}, {Loveday}, {Malanushenko}, {Malanushenko}, {Oravetz}, {Pan}, {Pier}, {Simmons}, {de Haas} and {York}}]{2011ApJS..194...45S}
\bibinfo{author}{{Shen}, Y.}, \bibinfo{author}{{Richards}, G.T.}, \bibinfo{author}{{Strauss}, M.A.}, \bibinfo{author}{{Hall}, P.B.}, \bibinfo{author}{{Schneider}, D.P.}, \bibinfo{author}{{Snedden}, S.}, \bibinfo{author}{{Bizyaev}, D.}, \bibinfo{author}{{Brewington}, H.}, \bibinfo{author}{{Brinkmann}, J.}, \bibinfo{author}{{Eisenstein}, D.J.}, \bibinfo{author}{{Frieman}, J.}, \bibinfo{author}{{Gray}, J.}, \bibinfo{author}{{Gunn}, J.E.}, \bibinfo{author}{{Hao}, L.}, \bibinfo{author}{{Harrington}, R.}, \bibinfo{author}{{Harvanek}, M.}, \bibinfo{author}{{Kent}, S.M.}, \bibinfo{author}{{Kleinman}, S.J.}, \bibinfo{author}{{Kniazev}, A.}, \bibinfo{author}{{Krzesinski}, J.}, \bibinfo{author}{{Long}, D.}, \bibinfo{author}{{Loveday}, J.}, \bibinfo{author}{{Malanushenko}, E.}, \bibinfo{author}{{Malanushenko}, V.}, \bibinfo{author}{{Oravetz}, D.}, \bibinfo{author}{{Pan}, K.}, \bibinfo{author}{{Pier}, J.R.}, \bibinfo{author}{{Simmons}, A.}, \bibinfo{author}{{de Haas}, E.}, \bibinfo{author}{{York}, D.G.},
  \bibinfo{year}{2011}.
\newblock \bibinfo{title}{{A Catalog of Quasar Properties from Sloan Digital Sky Survey Data Release 7}}.
\newblock \bibinfo{journal}{\apjs} \bibinfo{volume}{194}, \bibinfo{pages}{45}.
\newblock \DOIprefix\doi{10.1088/0067-0049/194/2/45}.
\bibitem[{{Sijacki} et~al.(2007){Sijacki}, {Springel}, {Di Matteo} and {Hernquist}}]{2007MNRAS.380..877S}
\bibinfo{author}{{Sijacki}, D.}, \bibinfo{author}{{Springel}, V.}, \bibinfo{author}{{Di Matteo}, T.}, \bibinfo{author}{{Hernquist}, L.}, \bibinfo{year}{2007}.
\newblock \bibinfo{title}{{A unified model for AGN feedback in cosmological simulations of structure formation}}.
\newblock \bibinfo{journal}{\mnras} \bibinfo{volume}{380}, \bibinfo{pages}{877--900}.
\newblock \DOIprefix\doi{10.1111/j.1365-2966.2007.12153.x}.
\bibitem[{{Soltan}(1982)}]{1982MNRAS.200..115S}
\bibinfo{author}{{Soltan}, A.}, \bibinfo{year}{1982}.
\newblock \bibinfo{title}{{Masses of quasars.}}
\newblock \bibinfo{journal}{\mnras} \bibinfo{volume}{200}, \bibinfo{pages}{115--122}.
\newblock \DOIprefix\doi{10.1093/mnras/200.1.115}.
\bibitem[{Vestergaard and Peterson(2006)}]{Vestergaard}
\bibinfo{author}{Vestergaard, M.}, \bibinfo{author}{Peterson, B.M.}, \bibinfo{year}{2006}.
\newblock \bibinfo{title}{Determining central black hole masses in distant active galaxies and quasars. ii. improved optical and uv scaling relationships}.
\newblock \bibinfo{journal}{The Astrophysical Journal} \bibinfo{volume}{641}, \bibinfo{pages}{689}.
\bibitem[{{Wu} et~al.(2015){Wu}, {Wang}, {Fan}, {Yi}, {Zuo}, {Bian}, {Jiang}, {McGreer}, {Wang}, {Yang}, {Yang}, {Thompson}, {Stern}, {Wang}, {Wang}, {Zhang}, {Yang}, {Wang} and {Kang}}]{2015Natur.518..512W}
\bibinfo{author}{{Wu}, X.B.}, \bibinfo{author}{{Wang}, F.}, \bibinfo{author}{{Fan}, X.}, \bibinfo{author}{{Yi}, W.M.}, \bibinfo{author}{{Zuo}, W.}, \bibinfo{author}{{Bian}, F.}, \bibinfo{author}{{Jiang}, L.}, \bibinfo{author}{{McGreer}, I.D.}, \bibinfo{author}{{Wang}, R.}, \bibinfo{author}{{Yang}, J.}, \bibinfo{author}{{Yang}, Q.}, \bibinfo{author}{{Thompson}, D.}, \bibinfo{author}{{Stern}, D.}, \bibinfo{author}{{Wang}, J.}, \bibinfo{author}{{Wang}, Y.}, \bibinfo{author}{{Zhang}, X.}, \bibinfo{author}{{Yang}, J.}, \bibinfo{author}{{Wang}, X.}, \bibinfo{author}{{Kang}, W.}, \bibinfo{year}{2015}.
\newblock \bibinfo{title}{{An ultraluminous quasar with a twelve-billion-solar-mass black hole at redshift 6.30}}.
\newblock \bibinfo{journal}{\nat} \bibinfo{volume}{518}, \bibinfo{pages}{512--515}.
\newblock \DOIprefix\doi{10.1038/nature14241}.
\bibitem[{{Yong} and {Ong}(2023)}]{2023MNRAS.524.3116Y}
\bibinfo{author}{{Yong}, S.Q.}, \bibinfo{author}{{Ong}, D.}, \bibinfo{year}{2023}.
\newblock \bibinfo{title}{{Conformal quantile regression for virial black hole mass estimation and uncertainty quantification}}.
\newblock \bibinfo{journal}{\mnras} \bibinfo{volume}{524}, \bibinfo{pages}{3116--3127}.
\newblock \DOIprefix\doi{10.1093/mnras/stad2190}.

\end{thebibliography}

\end{document}